\documentclass{article}

\input{tcilatex}
\begin{document}

\centerline{\large\bf SLOCC invariant and semi-invariants } 
\centerline{\large\bf for SLOCC classification of four-qubits
\footnote{The paper was supported by NSFC(Grants No. 60433050 and
60673034) and the basic research fund of Tsinghua university No:
JC2003043. }} 
\centerline{Dafa Li$^{a}$\footnote{email address:dli@math.tsinghua.edu.cn},
Xiangrong Li$^{b}$, Hongtao Huang$^{c}$, Xinxin Li$^{d}$ }

\centerline{$^a$ Dept of mathematical sciences, Tsinghua University, Beijing
100084 CHINA}

\centerline{$^b$ Department of Mathematics, University of California, Irvine, CA
92697-3875, USA}

\centerline{$^c$ Electrical Engineering and Computer Science Department} %
\centerline{ University of Michigan, Ann Arbor, MI 48109, USA}

\centerline{$^d$ Dept. of computer science, Wayne State University, Detroit, MI 48202,
USA}

Abstract

We show there are at least 28 distinct true SLOCC entanglement classes for
four-qubits by means of SLOCC invariant and semi-invariants and derive the
number of the degenerated SLOCC classes for $n$-qubits.



\section{Introduction}

Entanglement plays a key role in quantum computing and quantum information.
If two states can be obtained from each other by means of local operations
and classical communication (LOCC) with nonzero probability, we say that two
states have the same kind of entanglement\cite{Bennett}. It is well known
that a pure entangled state of two-qubits can be locally transformed into a
state $GHZ$. For multipartite systems, there are several inequivalent forms
of entanglement under asymptotic LOCC\cite{Bennett2}. Recently, many authors
have investigated the equivalence classes of three-qubit states specified
SLOCC (stochastic local operations and classical communication ) \cite{Acin}$%
-$\cite{LDF-PLA}. D\"{u}r et al. showed that for pure states of three-qubits
there are four different degenerated SLOCC entanglement classes and two
inequivalent true entanglement classes\cite{Dur}. A. Miyake discussed the
onionlike classification of SLOCC\ orbits and proposed the SLOCC equivalence
classes using the orbits\cite{Miyake03}. A.K. Rajagopal and R.W. Rendell
gave the conditions for the full separability and the biseparability\cite%
{Rajagopal}. In \cite{LDF-PLA} we gave the simple criteria for the complete
SLOCC\ classification for three-qubits and the criteria of a few classes for
four-qubits.

Verstraete et al.\cite{Moor2} considered the entanglement classes of
four-qubits under SLOCC and concluded that there exist nine families of
states corresponding to nine different ways of entanglement and claimed that
by determinant one SLOCC operations, a pure state of four qubits can be
transformed into one of the nine families of states. Clearly, this does not
say that each family is a SLOCC\ class. Then, how many SLOCC classes are
there for each family? what are representations? After investigating the
nine families by means of methods in this paper, we can say that each of the
first six families includes several SLOCC\ classes. We list SLOCC\ classes
of some families as follows. For Family $L_{ab_{3}}:a$ $(|0000\rangle
+|1111\rangle )+\frac{a+b}{2}($\ $|0101\rangle +|1010\rangle )+$\ $\frac{a-b%
}{2}($\ $|0110\rangle +|1001\rangle )+\frac{i}{\sqrt{2}}(|0001\rangle
+|0010\rangle +|0111\rangle +|1011\rangle )$, \ this consists of five true
SLOCC entanglement classes, which are $a=b=0$, i.e. class $|W\rangle $, $%
a=b\neq 0$, $a=-b\neq 0$, $a\neq \pm b\wedge 3a^{2}+b^{2}\neq 0$ and $a\neq
\pm b\wedge 3a^{2}+b^{2}=0$, respectively. For Family $L_{a_{4}}:a(|0000%
\rangle +|0101\rangle +|1010\rangle +|1111\rangle )+(i|0001\rangle
+|0110\rangle -i|1011\rangle )$, this includes two true SLOCC\ entanglement
classes: $a=0$ and $a\neq 0$. For Family $L_{a_{2}0_{3\oplus 1}}:$ $%
a(|0000\rangle +|1111\rangle )+|0011\rangle +|0101\rangle +|0110\rangle $,
it includes two SLOCC classes. When $a=0$, this becomes a product state of a
qubit state and 3-qubit $|W\rangle $. When $a\neq 0$, this is a true
entangled state.

In \cite{Lamata1}\cite{Lamata2}, the authors utilized the partition for
SLOCC\ classification of three-qubits and four-qubits. The idea for the
partition was early used to analyze the separability of $n$-qubits and
multipartite pure states in \cite{LDF}. In \cite{Lamata2}, the authors
declared there are eight Span classes and 16 true SLOCC entanglement classes
for four-qubits. By means of methods in this paper, we can illustrate how
many true SLOCC entanglement classes there are for each Span $\{.....\}$.
For example, for Span $\{O_{k}\Psi ,O_{k}\Psi \}$, canonical states are $%
|0000\rangle +|1100\rangle +a|0011\rangle +b|1111\rangle $ and $|0000\rangle
+|1100\rangle $ $+a|0001\rangle +a|0010\rangle +b|1101\rangle +b|1110\rangle 
$, where $a\neq b$ \cite{Lamata2}. It was not pointed out in \cite{Lamata2}
that what relation $a$ and $b$ satisfy to be a representation of SLOCC
class. It can be shown that for the former canonical state, $a=-b$ and $%
a\neq -b$ represent two true SLOCC classes, respectively while for the
latter canonical state, $ab=0$ and $ab\neq 0$ represent two true SLOCC
classes, respectively. We can also explain that each of Span $\{000,GHZ\}$,
Span $\{0_{i}\Psi ,0_{j}\Psi \}$ and Span $\{GHZ,W\}$ includes four true
SLOCC entanglement classes and Span $\{0_{k}\Psi ,GHZ\}$ includes more. Also
considering Span $\{000,000\}$, Span $\{000,0_{k}\Psi \}$ and Span $\{000,W\}
$, in total, the eight Spans $\{...\}$ in \cite{Lamata2}\ include much more
true SLOCC entanglement classes.

In this paper, we find the SLOCC invariant and semi-invariants for
four-qubits. Using the invariant and semi-invariants, we can determine if
two states belong to different SLOCC entanglement classes. We distinguish 28
distinct true entanglement classes, where permutations of the qubits are
allowed. This classification is not complete. It seems that there are more
true entanglement classes. The invariant and semi-invariants only require
simple arithmetic operations.

\section{\protect\bigskip SLOCC invariant and semi-invariants}

We discuss the system comprising four qubits A, B, C and D. The states of a
four-qubit system can be generally expressed as 
\begin{equation}
|\psi \rangle =\sum_{i=0}^{15}a_{i}|i\rangle .
\end{equation}

Two states $|\psi \rangle $ and $|\psi ^{\prime }\rangle $ are equivalent
under SLOCC if and only if there exist invertible local operators $\alpha
,\beta $, $\gamma $ and $\delta $\ such that

\begin{eqnarray}
|\psi \rangle =\alpha \otimes \beta \otimes \gamma \otimes \delta |\psi
^{\prime }\rangle ,  \label{eq1}
\end{eqnarray}

\noindent where the local operators $\alpha ,\beta ,$ $\gamma $ and $\delta $
can be expressed as $2\times 2$ invertible matrices

\begin{eqnarray}
\alpha =\left( 
\begin{array}{cc}
\alpha _{1} & \alpha _{2} \\ 
\alpha _{3} & \alpha _{4}%
\end{array}%
\right) ,~\beta =\left( 
\begin{array}{cc}
\beta _{1} & \beta _{2} \\ 
\beta _{3} & \beta _{4}%
\end{array}%
\right) ,~\gamma =\left( 
\begin{tabular}{cc}
$\ \gamma _{1}$ & $\ \gamma _{2}$ \\ 
$\ \gamma _{3}$ & $\ \gamma _{4}$%
\end{tabular}%
\right) ,\delta =\left( 
\begin{tabular}{cc}
$\ \delta _{1}$ & $\ \delta _{2}$ \\ 
$\ \delta _{3}$ & $\ \delta _{4}$%
\end{tabular}%
\right) .
\end{eqnarray}

\subsection{SLOCC\ invariant}

Let $|\psi ^{\prime }\rangle =\sum_{i=0}^{15}b_{i}|i\rangle $ in Eq. (\ref%
{eq1}). If $|\psi \rangle $ is SLOCC\ equivalent to $|\psi ^{\prime }\rangle 
$, then the following equation holds.

\begin{equation}
IV(\psi )=IV(\psi ^{\prime })\det (\alpha )\det (\beta )\det (\gamma )\det
(\delta ),  \label{IV0}
\end{equation}

where

\begin{equation}
IV(\psi
)=(a_{2}a_{13}-a_{3}a_{12})+(a_{4}a_{11}-a_{5}a_{10})-(a_{0}a_{15}-a_{1}a_{14})-(a_{6}a_{9}-a_{7}a_{8})
\end{equation}%
and 
\begin{equation}
IV(\psi ^{\prime
})=(b_{2}b_{13}-b_{3}b_{12})+(b_{4}b_{11}-b_{5}b_{10})-(b_{0}b_{15}-b_{1}b_{14})-(b_{6}b_{9}-b_{7}b_{8}).
\end{equation}

Eq. (\ref{IV0}) was by induction derived in \cite{LDF07}. We can also verify
Eq. (\ref{IV0}) as follows. By solving the matrix equation in Eq. (\ref{eq1}%
), we obtain amplitudes $a_{i}$ of state $|\psi \rangle $\ in Eq. (\ref{eq1}%
).\ Then substituting $a_{i}$ into $IV(\psi )$, we have Eq. (\ref{IV0}).

Notice that $IV(\psi )$ does not vary under determinant one SLOCC\
operations ($SL$-operations) or vanish under\textbf{\ }non-unit-determinant
SLOCC operations.

If $\psi $\ is $SL$-equivalent to $\psi ^{\prime }$, then $IV(\psi )=IV(\psi
^{\prime })$.\ Eq. (\ref{IV0}) implies that each SLOCC class has infinite $%
SL $-classes. For family $L_{a_{2}0_{3\oplus 1}}$ in \cite{Moor2}, let $%
|\psi ^{\prime }\rangle $ be a representative state of the family: $%
a(|0000\rangle +|1111\rangle )+|0011\rangle +|0101\rangle +|0110\rangle $.
Eq. (\ref{IV0}) becomes $IV(\psi )=-a^{2}\det (\alpha )\det (\beta )\det
(\gamma )\det (\delta )$. For $SL$-operations, $IV(\psi )=IV(\psi ^{\prime
})=-a^{2}$. It is clear that different $a^{\prime }$s values yield different 
$SL$-classes.\ Therefore there are infinite $SL$-classes when $a\neq 0$.
However, the infinite $SL$-classes belong to a true SLOCC entanglement class.

\subsection{Semi-invariants $F_{i}$}

Coffman defined the concurrence of three-qubits. We extend the definition of
the concurrence of three-qubits to the one of four-qubits as follows. For
state $|\psi \rangle $, we define $F_{i}(\psi )$ as follows. Notice that $%
F_{3}(\psi )$ to $F_{8}(\psi )$ can be obtained from $F_{1}(\psi )$ and $%
F_{2}(\psi )$ by permutations of the qubits.

$F(\psi )=4\sum_{i=1}^{10}\left\vert F_{i}(\psi )\right\vert $, where

$F_{1}(\psi
)=(a_{0}a_{7}-a_{2}a_{5}+(a_{1}a_{6}-a_{3}a_{4}))^{2}-4(a_{2}a_{4}-a_{0}a_{6})(a_{3}a_{5}-a_{1}a_{7}), 
$

$F_{2}(\psi
)=((a_{8}a_{15}-a_{11}a_{12})+(a_{9}a_{14}-a_{10}a_{13}))^{2}-4(a_{11}a_{13}-a_{9}a_{15})(a_{10}a_{12}-a_{8}a_{14}), 
$

$F_{3}(\psi
)=(a_{0}a_{11}-a_{2}a_{9}+a_{1}a_{10}-a_{3}a_{8})^{2}-4(a_{2}a_{8}-a_{0}a_{10})(a_{3}a_{9}-a_{1}a_{11}), 
$

$F_{4}(\psi
)=(a_{4}a_{15}-a_{6}a_{13}+a_{5}a_{14}-a_{7}a_{12})^{2}-4(a_{6}a_{12}-a_{4}a_{14})(a_{7}a_{13}-a_{5}a_{15}), 
$

$F_{5}(\psi
)=(a_{0}a_{13}-a_{4}a_{9}+a_{1}a_{12}-a_{5}a_{8})^{2}-4(a_{4}a_{8}-a_{0}a_{12})(a_{5}a_{9}-a_{1}a_{13}), 
$

$F_{6}(\psi
)=(a_{2}a_{15}-a_{6}a_{11}+a_{3}a_{14}-a_{7}a_{10})^{2}-4(a_{6}a_{10}-a_{2}a_{14})(a_{7}a_{11}-a_{3}a_{15}), 
$

$F_{7}(\psi
)=(a_{0}a_{14}-a_{4}a_{10}+a_{2}a_{12}-a_{6}a_{8})^{2}-4(a_{4}a_{8}-a_{0}a_{12})(a_{6}a_{10}-a_{2}a_{14}), 
$

$F_{8}(\psi
)=(a_{1}a_{15}-a_{5}a_{11}+a_{3}a_{13}-a_{7}a_{9})^{2}-4(a_{5}a_{9}-a_{1}a_{13})(a_{7}a_{11}-a_{3}a_{15}), 
$

$F_{9}(\psi
)=((a_{0}a_{15}-a_{2}a_{13})+(a_{1}a_{14}-a_{3}a_{12}))^{2}-4(a_{0}a_{14}-a_{2}a_{12})(a_{1}a_{15}-a_{3}a_{13}), 
$

$F_{10}(\psi
)=((a_{4}a_{11}-a_{7}a_{8})+(a_{5}a_{10}-a_{6}a_{9}))^{2}-4(a_{7}a_{9}-a_{5}a_{11})(a_{6}a_{8}-a_{4}a_{10})) 
$.

For state $|\psi ^{\prime }\rangle $, let $F_{i}(\psi ^{\prime })$\ be
obtained from $F_{i}(\psi )$ by replacing $a$ in $F_{i}(\psi )$ by $b$. Then
by induction we can show that $F_{i}$ have the following interesting
properties and the properties are called as semi-invariants.

In Eq. (\ref{eq1}), let $\alpha =I$, where $I$ is an identity. Thus, Eq. (%
\ref{eq1}) becomes 
\begin{equation}
|\psi \rangle =I\otimes \beta \otimes \gamma \otimes \delta |\psi ^{\prime
}\rangle .  \label{F12}
\end{equation}
Then we have the following. 
\begin{equation}
F_{i}(\psi )=F_{i}(\psi ^{\prime })\det^{2}(\beta )\det^{2}(\gamma
)\det^{2}(\delta ),i=1,2.  \label{semi-F12}
\end{equation}

Eq. (\ref{semi-F12}) can be verified as follows. We obtain amplitudes $a_{i}$
of state $|\psi \rangle $ by solving Eq. (\ref{F12}). Then substituting $%
a_{i}$ into $F_{i}(\psi )$, we derive Eq. (\ref{semi-F12}).

Also, in Eq. (\ref{eq1}), let $\beta =I$, then $F_{i}(\psi )=$\ $F_{i}(\psi
^{\prime })\det^{2}(\alpha )\det^{2}(\gamma )\det^{2}(\delta )$, $i=3,4$.

In Eq. (\ref{eq1}), let $\gamma =I$, then $F_{i}(\psi )=$\ $F_{i}(\psi
^{\prime })\det^{2}(\alpha )\det^{2}(\beta )\det^{2}(\delta )$, $i=5,6$.

In Eq. (\ref{eq1}), let $\delta =I$, then $F_{i}(\psi )=$\ $F_{i}(\psi
^{\prime })\det^{2}(\alpha )\det^{2}(\beta )\det^{2}(\gamma )$, $i=7,8$.

In Eq. (\ref{eq1}), let $\alpha =I$ and $\beta =I$, i.e. $|\psi \rangle
=I\otimes I\otimes \gamma \otimes \delta |\psi ^{\prime }\rangle $, then $%
F_{i}(\psi )=$\ $F_{i}(\psi ^{\prime })\det^{2}(\gamma )\det^{2}(\delta )$, $%
i=9,10$.

Next let $|\psi \rangle $ be SLOCC\ equivalent to $|\psi ^{\prime }\rangle $
in Eq. (\ref{eq1}). By solving the matrix equation in Eq. (\ref{eq1}), we
obtain amplitudes $a_{i}$ of state $|\psi \rangle $\ in Eq. (\ref{eq1}).
Then we can calculate $F_{i}$ of state $|\psi \rangle $, i.e., the $F_{i}$
of class $|\psi ^{\prime }\rangle $. We compute $F_{i}$ of all the
degenerated entanglement classes and 28 true entanglement classes. See
Tables 2.1, 2.2 and 5 and Appendix A. If $F_{i}$ do not vanish for some
classes, then we give the expressions of $F_{i}$ in Appendix A. We list the
properties of $F_{i}$ of the 28 true entanglement classes in Tables 2.1 and
2.2 and of $F_{i}$\ of all the degenerated entanglement classes in Table 5.
For the derivations of the properties of $F_{i}$, see Appendix A. We also
compute all the $F_{i}$ of the 28 true entanglement states, see Tables 3.1.
and 3.2.

\subsection{Semi-invariants $D_{1}$, $D_{2}$ and $D_{3}$}

In \cite{LDF-PLA}, we computed the following expressions of $D_{1}$, $D_{2}$
and $D_{3}$ for $|GHZ\rangle $, $|W\rangle .$ Using the $D_{1}$, $D_{2}$ and 
$D_{3}$, we found a true entanglement state $|C_{4}\rangle $ which is
distinct from $|GHZ\rangle $, $|W\rangle $ and $|\phi _{4}\rangle $\cite%
{Briegel}. We define $D_{i}(\psi )$ for state $\psi $ as follows.

\begin{equation}
D_{1}(\psi
)=(a_{1}a_{4}-a_{0}a_{5})(a_{11}a_{14}-a_{10}a_{15})-(a_{3}a_{6}-a_{2}a_{7})(a_{9}a_{12}-a_{8}a_{13}),
\end{equation}

\begin{equation}
D_{2}(\psi
)=(a_{4}a_{7}-a_{5}a_{6})(a_{8}a_{11}-a_{9}a_{10})-(a_{0}a_{3}-a_{1}a_{2})(a_{12}a_{15}-a_{13}a_{14}),
\end{equation}

\begin{equation}
D_{3}(\psi
)=(a_{3}a_{5}-a_{1}a_{7})(a_{10}a_{12}-a_{8}a_{14})-(a_{2}a_{4}-a_{0}a_{6})(a_{11}a_{13}-a_{9}a_{15}).
\end{equation}

Then by induction we can demonstrate that $D_{i}(\psi )$ have the following
interesting properties, which are also called as semi-invariants. The
properties can also be verified by substituting the amplitudes $a_{i}$ of
state $|\psi \rangle $\ in Eq. (\ref{eq1}) into $D_{i}(\psi )$ $i=1,2$ and $%
3 $. For state $|\psi ^{\prime }\rangle $, let $D_{i}(\psi ^{\prime })$ be
be obtained from $D_{i}(\psi )$ by replacing $a$ in $D_{i}(\psi )$ by $b$.
Then

In Eq. (\ref{eq1}), let $\alpha =I$ and $\gamma =I$, i.e. $|\psi \rangle
=I\otimes \beta \otimes I\otimes \delta |\psi ^{\prime }\rangle $, then $%
D_{1}(\psi )=$\ $D_{1}(\psi ^{\prime })\det^{2}(\beta )\det^{2}(\delta )$.

In Eq. (\ref{eq1}), let $\alpha =I$ and $\beta =I$, then $D_{2}(\psi )=$\ $%
D_{2}(\psi ^{\prime })\det^{2}(\gamma )\det^{2}(\delta )$.

In Eq. (\ref{eq1}), let $\alpha =I$ and $\delta =I$, then $D_{3}(\psi )=$\ $%
D_{3}(\psi ^{\prime })\det^{2}(\beta )\det^{2}(\gamma )$.

Next let $|\psi \rangle $ be SLOCC\ equivalent to $|\psi ^{\prime }\rangle $
in Eq. (\ref{eq1}).\ By solving the matrix equation in Eq. (\ref{eq1}), we
obtain amplitudes $a_{i}$ of state $|\psi \rangle $\ in Eq. (\ref{eq1}).
Then we can calculate $D_{i}$ of state $|\psi \rangle $, i.e., the $D_{i}$
of class $|\psi ^{\prime }\rangle $.\ We compute the values of $D_{1},D_{2}$
and $D_{3}$ of the degenerated SLOCC equivalence classes (see Table 4). We
give the values of $D_{1},D_{2}$ and $D_{3}$\ of the 28 true entanglement
classes in Tables 1.1 and 1.2 and of the 28 true entanglement states in
Tables 3.1 and 3.2. We also calculate the values of $D_{1},D_{2}$ and $D_{3}$%
\ of states $|GHZ\rangle _{12}\otimes |GHZ\rangle _{34}$, $|GHZ\rangle
_{13}\otimes |GHZ\rangle _{24}$ and $|GHZ\rangle _{14}\otimes |GHZ\rangle
_{23}$, see Table 6. If $D_{i}=0$ for some $i$ and for some class in Tables
1.1 and 1.2 and 4, then it implies that $D_{i}=0$ for the $i$ and for all
the states of the class. If $D_{i}$ is \textquotedblleft
opt\textquotedblright\ for some $i$ and for some class in Tables 1.1 and 1.2
and 4, then it means that $D_{i}=0$ for the $i$ and for some states of the
class while for other states of the class $D_{i}\neq 0$. If $D_{i}$ is
\textquotedblleft opt\textquotedblright , then we give the expression of $%
D_{i}$ in Appendix A. For example, for class $|\kappa _{4}\rangle $ in Table
1.1, $D_{1}$ is \textquotedblleft opt\textquotedblright , $D_{2}$ is
\textquotedblleft opt\textquotedblright\ and $D_{3}=0$. It says that for
some state of class $|\kappa _{4}\rangle $ in Table 1.1, $D_{1}\neq 0$ and $%
D_{2}\neq 0$ but $D_{3}=0$ for every state of class $|\kappa _{4}\rangle $.
However, for state $|\kappa _{4}\rangle $ in Table 3.1, $D_{i}$ $=0$, where $%
i=1,2$ and $3$,

\section{The invariant, semi-invariants for SLOCC classification}

\subsection{The representatives of true entanglement classes}

It is well known that states $|GHZ\rangle $, $|W\rangle $, $|\phi
_{4}\rangle $ and $|C_{4}\rangle $ are the representatives of disjoint true
entanglement classes of four-qubits. Utilizing the SLOCC invariant $IV$ and
the semi-invariants $F_{i}$ and $D_{i}$ of four-qubits, we find 28 distinct
true entanglement classes. The representatives of the classes are listed
below.\ 

1. From the construction of $|\phi _{4}\rangle $, we do the following
computation tests.\ From all the $15$ true entanglement states: $(|0\rangle
+|i\rangle +|j\rangle -|15\rangle )/2$, where $|i\rangle ,|j\rangle \in
\{|3\rangle ,|5\rangle ,|6\rangle ,|9\rangle ,|10\rangle ,|12\rangle \}$
which is obtained from $|C_{4}\rangle $, we find the representatives of 7
different true entanglement classes. They are $|GHZ\rangle ,|\phi
_{4}\rangle ,|\psi _{4}\rangle $, $|\mu _{4}\rangle $, $|\kappa _{4}\rangle $%
, $|E_{4}\rangle $ and $|L_{4}\rangle $.

2. From \cite{Moor2}, we consider the states of the following forms: $%
(|0\rangle +|i\rangle +|j\rangle +|k\rangle +|l\rangle +|15\rangle )/\sqrt{6}
$, where $|i\rangle ,|j\rangle ,|k\rangle ,|l\rangle \in \{|3\rangle
,|5\rangle ,|6\rangle ,|9\rangle ,|10\rangle ,|12\rangle $. There are 15
true entanglement states, of which seven are chosen as the representatives
of different true entanglement classes. They are $|C_{4}\rangle ,|\pi
_{4}\rangle ,|\sigma _{4}\rangle ,|\rho _{4}\rangle ,|\xi _{4}\rangle
,|\epsilon _{4}\rangle $ and $|\theta _{4}\rangle $.

3. From the 15 true entanglement states: $(|0\rangle +|i\rangle +|j\rangle
+|k\rangle +|l\rangle -|15\rangle )/\sqrt{6}$, where $|i\rangle ,|j\rangle
,|k\rangle ,|l\rangle \in \{|3\rangle ,|5\rangle ,|6\rangle ,|9\rangle
,|10\rangle ,|12\rangle $, we choose $|\chi _{4}\rangle $ as a
representative.

4. Consider the true entanglement states of the following forms:\ $|3\rangle
+|x\rangle +|12\rangle $, where $|x\rangle \in \{|5\rangle ,|6\rangle
,|9\rangle ,|10\rangle \}$; \ $|5\rangle +|x\rangle +|10\rangle $, where $%
|x\rangle \in \{|3\rangle ,|6\rangle ,|9\rangle ,|12\rangle \}$; \ $%
|6\rangle +|x\rangle +|9\rangle $, where $|x\rangle \in \{|3\rangle
,|5\rangle ,|10\rangle ,|12\rangle \}$. Notice that $|3\rangle $ and $%
|12\rangle $, $|5\rangle $ and $|10\rangle |$ and $6\rangle $ and $|9\rangle 
$ are dual, respectively. From the 12 true entanglement states, we find
three inequivalent true entanglement states. They are $|H_{4}\rangle
,|\lambda _{4}\rangle $ and $|M_{4}\rangle $.

5. The classes whose representatives have 3 product terms

Let $S_{1}=(001)^{T}$, $S_{2}=(010)^{T}$, $S_{3}=(100)^{T}$, $%
V_{1}=(011)^{T} $, $V_{2}=(101)^{T}$, $V_{3}=(110)^{T}$. Consider the
permutations: $S_{i}S_{j}V_{i}V_{j}$. For example, $S_{1}S_{2}V_{1}V_{2}$ is
considered as a matrix $\left( 
\begin{array}{cccc}
0 & 0 & 0 & 1 \\ 
0 & 1 & 1 & 0 \\ 
1 & 0 & 1 & 1%
\end{array}%
\right) $. Each row of the matrix is considered a basic state of
four-qubits. Thus, the matrix can be considered a state $(|1\rangle
+|6\rangle +|11\rangle )/\sqrt{3}$. From the permutations, we find
representatives: $|\varphi _{4}\rangle $, $|\tau _{4}\rangle $, $|\vartheta
_{4}\rangle $, $|\varrho _{4}\rangle $, $|\iota _{4}\rangle $, $|\varsigma
_{4}\rangle $.

6. $|\omega _{4}\rangle $ is $L_{0_{5\oplus 3}}$ in \cite{Moor2}. From
states of the forms $(|x\rangle +|5\rangle +|y\rangle +|z\rangle )/2$, where 
$x+5+y+z=27$, we choose $|\upsilon _{4}\rangle $, $|\varpi _{4}\rangle $ and 
$|\omega _{4}\rangle $ as representatives. \ \ 

7. Up to permutations of the qubits, each one of the following groups is
considered as one true entanglement class. However, we don't show that
different groups can not be obtained up to permutations of the qubits.

We list the 28 true entanglement classes as follows.

1. $|GHZ\rangle =(|0\rangle +$ $|15\rangle )/\sqrt{2},$

2. $|W\rangle =(|1\rangle +|2\rangle +|4\rangle +|8\rangle )/2,$

3. $|C_{4}\rangle =(|3\rangle +|5\rangle +|6\rangle +|9\rangle +|10\rangle
+|12)/\sqrt{6}$,

4. $\ $

$|\kappa _{4}\rangle =(|0\rangle +|3\rangle +|10\rangle -|15\rangle )/2,$

$|E_{4}\rangle =(|0\rangle +|5\rangle +|9\rangle -|15\rangle )/2,$

$|L_{4}\rangle =(|0\rangle +|3\rangle +|9\rangle -|15\rangle )/2,$

5.

$|H_{4}\rangle =(|3\rangle +|6\rangle +|12)/\sqrt{3},$

$|\lambda _{4}\rangle =(|5\rangle +|6\rangle +|10)/\sqrt{3},$

$|M_{4}\rangle =(|3\rangle +|5\rangle +|12)/\sqrt{3},$

6.

$|\pi _{4}\rangle =(|0\rangle +|3\rangle +|5\rangle +|6\rangle +|10\rangle
+|15\rangle )/\sqrt{6},$

$|\theta _{4}\rangle =(|0\rangle +|5\rangle +|6\rangle +|10\rangle
+|12\rangle +|15\rangle )/\sqrt{6},$

$|\sigma _{4}\rangle =(|0\rangle +|3\rangle +|9\rangle +|10\rangle
+|12\rangle +|15\rangle )/\sqrt{6},$

$|\rho _{4}\rangle =(|0\rangle +|3\rangle +|6\rangle +|10\rangle +|12\rangle
+|15\rangle )/\sqrt{6},$

$|\xi _{4}\rangle =(|0\rangle +|6\rangle +|9\rangle +|10\rangle +|12\rangle
+|15\rangle )/\sqrt{6},$

$|\epsilon _{4}\rangle =(|0\rangle +|3\rangle +|6\rangle +|9\rangle
+|10\rangle +|15\rangle )/\sqrt{6},$

7.

$|\chi _{4}\rangle =(|0\rangle +|3\rangle +|6\rangle +|10\rangle +|12\rangle
-|15\rangle )/\sqrt{6},$

8.

$|\psi _{4}\rangle =$ $(|0\rangle +|5\rangle +|10\rangle -|15\rangle )/2,$

$|\phi _{4}\rangle =$ $(|0\rangle +|3\rangle +|12\rangle -|15\rangle )/2,$

$|\mu _{4}\rangle =$ $(|0\rangle +|6\rangle +|9\rangle -|15\rangle )/2,$

9.

$|\varphi _{4}\rangle =(|1\rangle +|6\rangle +|11\rangle )/\sqrt{3},$

$|\vartheta _{4}\rangle =(|2\rangle +|5\rangle +|11\rangle )/\sqrt{3},$

$|\tau _{4}\rangle =(|1\rangle +|7\rangle +|10\rangle )/\sqrt{3},$

$|\varrho _{4}\rangle =(|2\rangle +|7\rangle +|9\rangle )/\sqrt{3},$

10.

$|\zeta _{4}\rangle =(|0\rangle +|11\rangle +|12\rangle )/\sqrt{3},$

$|\iota _{4}\rangle =(|0\rangle +|3\rangle +|13\rangle )/\sqrt{3},$

11.

$|\upsilon _{4}\rangle =(|2\rangle +|5\rangle +|9\rangle +|11\rangle /2,$

12.

$|\omega _{4}\rangle =(|0\rangle +|5\rangle +|8\rangle +|14\rangle )/2,$

13.

$|\varpi _{4}\rangle =(|2\rangle +|5\rangle +|8\rangle +|12\rangle )/2$

\subsection{The sufficient conditions for a true entanglement state}

From Tables 1, 2, 4 and 5, it is not difficult to see that a state is a true
entanglement state if the state satisfies one of the following conditions.

(1). $IV=0$ and $D_{i}\neq 0$, where $i=1,2$ or $3$,

(2). $IV\neq 0$ and $F_{i}\neq 0$, where $i=1,2,3,4,5,6,7$ or $8$,

(3). $IV\neq 0$ and $D_{i}\neq 0$ and $D_{j}\neq 0$.

\section{At least 28 distinct true entanglement classes}

\subsection{Degenerated entanglement classes}

The authors in \cite{Lamata1} gave an upper bound for the number of
degenerate $(N+1)$-entanglement classes in terms of the number of $N$%
-partite entanglement classes. In this paper, we give an exact recursive
formula\textbf{\ }for the number of degenerate entanglement classes of $n$%
-qubits, see Appendix B. By the recursive formula, for five-qubits, there
are $5\ast t(4)+66$ distinct degenerated SLOCC entanglement classes, where $%
t(4)$ is the number of the true SLOCC entanglement classes for four-qubits.\
We only use combinatory analysis to derive the recursive formula. The
authors in \cite{Lamata2} declared there are 16 true entanglement SLOCC
classes for four-qubits and at most 170 degenerate entanglement classes for
five-qubits. If so, by our recursive formula there would be 146 degenerated
SLOCC entanglement classes for five-qubits. However, in this paper, we
report there are at least 28 true entanglement SLOCC classes for
four-qubits. Thus, by the recursive formula there are at least 206
degenerated SLOCC classes for five-qubits. From the recursive formula, we
know that the most of the degenerated entanglement classes of $n$-qubits are
the $(n-1)$-qubit true entanglement with a separate qubit like $%
A-(BC....Z)_{n-1}$, where $(BC....Z)_{n-1}$ is truly entangled.

For four-qubits, by computing, we obtain the SLOCC invariant, the
semi-invariants $F_{i}$ and $D_{i}$\ of all the degenerated entanglement
classes. See Tables 4, 5 and 6. For example, value $F$ of three-qubit $GHZ$
entanglement accompanied with a separable qubit does not vanish. This proof
is given as follows. Other proofs are omitted.

By the definition of $F$ and section 3.1 of \cite{LDF-PLA}, it is easy to
obtain this result.

For class $|GHZ\rangle _{ABC}\otimes (s|0\rangle +t|1\rangle )_{D}$, \ by
section 3.1 of \cite{LDF-PLA}, $%
(a_{0}a_{14}-a_{4}a_{10}+a_{2}a_{12}-a_{6}a_{8})^{2}\neq
4(a_{4}a_{8}-a_{0}a_{12})(a_{6}a_{10}-a_{2}a_{14})$ or

$(a_{1}a_{15}-a_{5}a_{11}+a_{3}a_{13}-a_{7}a_{9})^{2}\neq
4(a_{5}a_{9}-a_{1}a_{13})(a_{7}a_{11}-a_{3}a_{15})$, and other $F_{i}$
vanish.

For class $|GHZ\rangle _{ABD}\otimes (s|0\rangle +t|1\rangle )_{C}$,

\ $(a_{0}a_{13}-a_{4}a_{9}+a_{1}a_{12}-a_{5}a_{8})^{2}\neq
4(a_{4}a_{8}-a_{0}a_{12})(a_{5}a_{9}-a_{1}a_{13})$

or $(a_{2}a_{15}-a_{6}a_{11}+a_{3}a_{14}-a_{7}a_{10})^{2}\neq
4(a_{6}a_{10}-a_{2}a_{14})(a_{7}a_{11}-a_{3}a_{15})$,

and other $F_{i}$ vanish.

For\ class $|GHZ\rangle _{ACD}\otimes (s|0\rangle +t|1\rangle )_{B}$ ,

$(a_{0}a_{11}-a_{2}a_{9}+a_{1}a_{10}-a_{3}a_{8})^{2}\neq
4(a_{2}a_{8}-a_{0}a_{10})(a_{3}a_{9}-a_{1}a_{11})$

or $(a_{4}a_{15}-a_{6}a_{13}+a_{5}a_{14}-a_{7}a_{12})^{2}\neq
4(a_{6}a_{12}-a_{4}a_{14})(a_{7}a_{13}-a_{5}a_{15})$.

For class $(s|0\rangle +t|1\rangle )_{A}\otimes $ $|GHZ\rangle _{BCD}$,

$(a_{0}a_{7}-a_{2}a_{5}+(a_{1}a_{6}-a_{3}a_{4}))^{2}\neq
4(a_{2}a_{4}-a_{0}a_{6})(a_{3}a_{5}-a_{1}a_{7})$

or $((a_{8}a_{15}-a_{11}a_{12})+(a_{9}a_{14}-a_{10}a_{13}))^{2}\neq
4(a_{11}a_{13}-a_{9}a_{15})(a_{10}a_{12}-a_{8}a_{14})$.

\subsection{The 28 classes in Tables 1.1 and 1.2 are true entanglement
classes.}

It is known that classes $|GHZ\rangle $, $|W\rangle $, $|\phi _{4}\rangle $%
\cite{Briegel} and $|C_{4}\rangle $ \cite{LDF-PLA} are inequivalent true
entanglement classes.

\bigskip Part 1. The classes in Table 1.1 are true entanglement classes.

Since for the classes in Table 1.1, $IV\neq 0$ and $F>0$, so we only need to
show that the classes in Table 1.1 are distinct from the degenerated
entanglement classes $|GHZ\rangle _{13}\otimes |GHZ\rangle _{24}$ and $%
|GHZ\rangle _{14}\otimes |GHZ\rangle _{23}$. However, $F_{i}$ in Tables 3.1
do not satisfy the properties of $F_{i}$ of classes $|GHZ\rangle
_{13}\otimes |GHZ\rangle _{24}$ or $|GHZ\rangle _{14}\otimes |GHZ\rangle
_{23}$\ in Table 5. Hence, the classes in Table 1.1 are true entanglement
classes.

Part 2. The classes in Table 1.2 are true entanglement classes.

Since $IV=0$ for classes in Table 1.2, the classes in Table 1.2 are distinct
from the degenerated entanglement classes $|GHZ\rangle _{12}\otimes
|GHZ\rangle _{34}$, $|GHZ\rangle _{13}\otimes |GHZ\rangle _{24}$ and $%
|GHZ\rangle _{14}\otimes |GHZ\rangle _{23}$. For some states of classes $%
|\chi _{4}\rangle $, $|\upsilon _{4}\rangle $, $|\varpi _{4}\rangle $, $%
|\psi _{4}\rangle $, $|\phi _{4}\rangle $, $|\mu _{4}\rangle $, $|\varphi
_{4}\rangle $, $|\varsigma _{4}\rangle $ and $|\vartheta _{4}\rangle $,
always $D_{i}\neq 0$ for some $i$, see Appendix A. However, all the classes
in Table 4 except for $|GHZ\rangle _{12}\otimes |GHZ\rangle _{34}$, $%
|GHZ\rangle _{13}\otimes |GHZ\rangle _{24}$ and $|GHZ\rangle _{14}\otimes
|GHZ\rangle _{23}$, require $D_{i}=0$, where $i=1,2,$ and $3$, see Table 4.
So classes $|\chi _{4}\rangle $, $|\upsilon _{4}\rangle $, $|\varpi
_{4}\rangle $, $|\psi _{4}\rangle $, $|\phi _{4}\rangle $, $|\mu _{4}\rangle 
$, $|\varphi _{4}\rangle $, $|\varsigma _{4}\rangle $ and $|\vartheta
_{4}\rangle $ are not degenerated entanglement classes. The classes $|\tau
_{4}\rangle $, $|\varrho _{4}\rangle $, $|\iota _{4}\rangle $, $|\omega
_{4}\rangle $ in Table 1.2 are not degenerated entanglement classes because
the properties of $F_{i}$ of classes $|\tau _{4}\rangle $, $|\varrho
_{4}\rangle $, $|\iota _{4}\rangle $, $|\omega _{4}\rangle $\ in Table 2.2
do not satisfy the conditions of $F_{i}$ in Table 5.

\subsection{The 28 classes in Tables 1.1 and 1.2 are distinct each other.}

Clearly, the classes in Table 1.1 are distinct from the ones in Table 1.2
because the values of $IV$\ of all the classes in Table 1.2 are zero while\
the values of $IV$ of all the classes in Table 1.1 are not zero.

Part 1. Let us show that the classes in Table 1.1 are distinct each other.

For class $|GHZ\rangle $, $D_{i}$ $=0$, where $i=1,2$ and $3$,$\ $see Table
1.1. However, always $D_{i}\neq 0$ for some $i$ and for some states of other
classes in Table 1.1. Consequently, class $|GHZ\rangle $ is distinct from
other classes in Table 1.1. For state $|C_{4}\rangle $, $D_{i}$ $\neq 0$,
where $i=1,2$ and $3$, see Table 3.1. It is easy to see from Table 1.1 that
for the other classes, always $D_{i}=0$ for some $i$. For example, $D_{3}=0$
for class $|\kappa _{4}\rangle $, see Table 1.1. It implies that $D_{3}=0$
for every state of class $|\kappa _{4}\rangle $. Therefore, class $%
|C_{4}\rangle $ is different from other classes in Table 1.1.

For some states of class $|\kappa _{4}\rangle $, $D_{1}\neq 0$ and $%
D_{2}\neq 0$, see the case for class $|\kappa _{4}\rangle $ in Appendix A,
and for every state of class $|\kappa _{4}\rangle $, $D_{3}=0$. Therefore
class $|\kappa _{4}\rangle $ is different from the last 11 classes in Table
1.1. As well, classes $|E_{4}\rangle $ and $|L_{4}\rangle $ are different
each other and from the last 9 classes in Table 1.1.

Let us demonstrate that class $|H_{4}\rangle $ is different from classes $%
|\pi _{4}\rangle $ and $|\theta _{4}\rangle $. For state $|H_{4}\rangle $, $%
F_{9}\neq 0$ and $F_{i}=0$ when $i\neq 9$, see Table 3.1. Thus, state $%
|H_{4}\rangle $ does not satisfy the conditions of $F_{i}$ of classes $|\pi
_{4}\rangle $ or $|\theta _{4}\rangle $, see Table 2.1. Therefore, class $%
|H_{4}\rangle $ is different from classes $|\pi _{4}\rangle $ and $|\theta
_{4}\rangle $. For some states of class $|H_{4}\rangle $, $D_{1}\neq 0$ and
for every state of class $|H_{4}\rangle $, $D_{2}=0$ and $D_{3}=0$,
therefore class $|H_{4}\rangle $ is different from $|\lambda _{4}\rangle $
and $|M_{4}\rangle $ and the last 4 classes in Table 1.1. As well, $|\lambda
_{4}\rangle $ and $|M_{4}\rangle $\ are different each other and from the
last 6 classes in Table 1.1. \ 

From Table 2.1, it is easy to see that for the last six classes: $|\pi
_{4}\rangle $, $|\theta _{4}\rangle $, $|\sigma _{4}\rangle $, $|\rho
_{4}\rangle $, $|\xi _{4}\rangle $ and $|\epsilon _{4}\rangle $, the
properties of $F_{i}$ are disjoint, hence they are distinct each other.

Part 2. We argue that the classes in Table 1.2 are distinct each other.

Since $F=0$ for class $|W\rangle $ and $F\neq 0$ for other classes in Table
1.2, class $|W\rangle $ is distinct from other classes in Table 1.2. For
some states of class $|\chi _{4}\rangle $, $D_{i}$ $\neq 0$, where $i=1,2$
and $3$, see the cases for class $|\chi _{4}\rangle $ in Appendix A. As
discussed for class $|C_{4}\rangle $ in Part 1, class $|\chi _{4}\rangle $
is distinct from other classes in Table 1.2.

For some states of class $|\upsilon _{4}\rangle $, $D_{2}\neq 0$ and $%
D_{3}\neq 0$, see the cases for class $|\upsilon _{4}\rangle $ in Appendix
A, and for every state of class $|\upsilon _{4}\rangle $, $D_{1}=0$, see
Table 1.2. Therefore, class $|\upsilon _{4}\rangle $ is different from other
classes in Table 1.2. We omit the similar discussions which can be found in
Part 1. We need to argue that $|\psi _{4}\rangle $ and $|\varphi _{4}\rangle 
$ are different each other. For state $|\psi _{4}\rangle $, $F_{1}=0$ and $%
F_{2}=0$, see Table 3.2. This conflicts that $\left\vert F_{1}\right\vert
+\left\vert F_{2}\right\vert \neq 0$ for class $|\varphi _{4}\rangle $, see
Table 2.2. This is done. As well, $|\phi _{4}\rangle $ and $|\varsigma
_{4}\rangle $ are different each other, so are $|\mu _{4}\rangle $ and $%
|\vartheta _{4}\rangle $. What remains is to explain that $|\tau _{4}\rangle 
$, $|\varrho _{4}\rangle $, $|\iota _{4}\rangle $ and $|\omega _{4}\rangle $
are distinct each other. This is obvious because the properties of their $%
F_{i}$ are disjoint, see Table 2.2.

Conjecture:

There should be many many true SLOCC entanglement classes. We can show $%
\frac{1}{\sqrt{6}}(\sqrt{2}|15\rangle +|8\rangle +|4\rangle +|2\rangle
+|1\rangle )$ in \cite{Osterloh} is a true entanglement state which does not
belong to the 28 classes because the state has the following properties.

$IV=0$, $D_{1}$: opt, $D_{2}$: opt, $D_{3}$: opt, ${\small |F}_{1}{\small %
|+|F}_{2}{\small |\neq 0,}$ ${\small |F}_{3}{\small |+|F}_{4}{\small |\neq
0,|F_{5}|+|F_{6}|\neq 0,|F_{7}|+|F_{8}|\neq 0}$.

Table 1.1.

The SLOCC invariants of true entanglement classes

\noindent 
\begin{tabular}{cccccc}
{\small classes} & ${\small F}$ & ${\small D}_{1}$ & ${\small D}_{2}$ & $%
{\small D}_{3}$ & ${\small IV}$ \\ 
\ $|GHZ\rangle $ & $>0$ & $=0$ & $=0$ & $=0$ & $\neq 0$ \\ 
\ $|C_{4}\rangle $ & $>0$ & opt & opt & opt & $\neq 0$ \\ 
$|\kappa _{4}\rangle $ & $>0$ & opt & opt & $=0$ & $\neq 0$ \\ 
$|E_{4}\rangle $ & $>0$ & opt & $=0$ & opt & $\neq 0$ \\ 
$|L_{4}\rangle $ & $>0$ & $=0$ & opt & opt & $\neq 0$ \\ 
$|H_{4}\rangle $ & $>0$ & opt & $=0$ & $=0$ & $\neq 0$ \\ 
\ $|\lambda _{4}\rangle $ & $>0$ & $=0$ & opt & $=0$ & $\neq 0$ \\ 
$|M_{4}\rangle $ & $>0$ & $=0$ & $=0$ & opt & $\neq 0$ \\ 
\ $|\pi _{4}\rangle $ & $>0$ & opt & $=0$ & $=0$ & $\neq 0$ \\ 
$|\theta _{4}\rangle $ & $>0$ & opt & $=0$ & $=0$ & $\neq 0$ \\ 
$|\sigma _{4}\rangle $ & $>0$ & $=0$ & opt & $=0$ & $\neq 0$ \\ 
$|\rho _{4}\rangle $ & $>0$ & $=0$ & opt & $=0$ & $\neq 0$ \\ 
$|\xi _{4}\rangle $ & $>0$ & $=0$ & $=0$ & opt & $\neq 0$ \\ 
$|\epsilon _{4}\rangle $ & $>0$ & $=0$ & $=0$ & opt & $\neq 0$%
\end{tabular}

\textquotedblleft opt\textquotedblright\ means that $D_{i}$ of some states
of the class\ are zero while $D_{i}$ of other states are not zero.

Table 1.2 The SLOCC invariants of true entanglement classes

\begin{tabular}{cccccc}
{\small classes} & ${\small F}$ & ${\small D}_{1}$ & ${\small D}_{2}$ & $%
{\small D}_{3}$ & ${\small IV}$ \\ 
\ $|W\rangle $ & $=0$ & $=0$ & $=0$ & $=0$ & $=0$ \\ 
$|\chi _{4}\rangle $ & $>0$ & opt & opt & opt & $=0$ \\ 
$|\upsilon _{4}\rangle $ & $>0$ & $=0$ & opt & opt & $=0$ \\ 
$|\varpi _{4}\rangle $ & $>0$ & opt & $=0$ & opt & $=0$ \\ 
\ $|\psi _{4}\rangle $ & $>0$ & opt & $=0$ & $=0$ & $=0$ \\ 
\ $|\phi _{4}\rangle $ & $>0$ & $=0$ & opt & $=0$ & $=0$ \\ 
\ $|\mu _{4}\rangle $ & $>0$ & $=0$ & $=0$ & opt & $=0$ \\ 
$|\varphi _{4}\rangle $ & $>0$ & opt & $=0$ & $=0$ & $=0$ \\ 
$|\zeta _{4}\rangle $ & $>0$ & $=0$ & opt & $=0$ & $=0$ \\ 
$|\vartheta _{4}\rangle $ & $>0$ & $=0$ & $=0$ & opt & $=0$ \\ 
$|\tau _{4}\rangle $ & $>0$ & $=0$ & $=0$ & $=0$ & $=0$ \\ 
$|\varrho _{4}\rangle $ & $>0$ & $=0$ & $=0$ & $=0$ & $=0$ \\ 
$|\iota _{4}\rangle $ & $>0$ & $=0$ & $=0$ & $=0$ & $=0$ \\ 
$|\omega _{4}\rangle $ & $>0$ & $=0$ & $=0$ & $=0$ & $=0$ \\ 
&  &  &  &  & 
\end{tabular}

Table 2.1.

The properties of $F_{i}$ for true entanglement classes

\noindent 
\begin{tabular}{ccc}
{\small classes} & ${\small F}_{i}$ & ${\small F}_{i}=0,{\small i=}$ \\ 
{\small \ }$|GHZ\rangle $ & $\#0$ &  \\ 
{\small \ }$|C_{4}\rangle $ & ${\small \#4}$ &  \\ 
$|\kappa _{4}\rangle ,|E_{4}\rangle ,|L_{4}\rangle ,|H_{4}\rangle ,${\small %
\ }$|\lambda _{4}\rangle ,|M_{4}\rangle $ & $\#0$ &  \\ 
{\small \ }$|\pi _{4}\rangle $ & ${\small |F}_{1}{\small |+|F}_{2}{\small %
|\neq 0,|F_{5}|+|F_{6}|\neq 0,\#2}$ & ${\small 3,4,7,8}$ \\ 
$|\theta _{4}\rangle $ & ${\small |F}_{3}{\small |+|F}_{4}{\small |\neq 0,|F}%
_{7}{\small |+|F}_{8}{\small |\neq 0,\#3}$ & ${\small 1,2,5,6}$ \\ 
$|\sigma _{4}\rangle $ & ${\small |F}_{1}{\small |+|F}_{2}{\small |\neq 0,|F}%
_{3}{\small |+|F}_{4}{\small |\neq 0,\#1}$ & ${\small 5,6,7,8}$ \\ 
$|\rho _{4}\rangle $ & ${\small |F}_{5}{\small |+|F}_{6}{\small |\neq 0,|F}%
_{7}{\small |+|F}_{8}{\small |\neq 0}$ & ${\small 1,2,3,4,9,10}$ \\ 
$|\xi _{4}\rangle $ & ${\small |F}_{1}{\small |+|F}_{2}{\small |\neq
0,|F_{7}|+|F_{8}|\neq 0,\#2}$ & ${\small 3,4,5,6}$ \\ 
$|\epsilon _{4}\rangle $ & ${\small |F}_{3}{\small |+|F}_{4}{\small |\neq
0,|F_{5}|+|F_{6}|\neq 0,\#3}$ & ${\small 1,2,7,8}$%
\end{tabular}

$\#0:$ If ${\small F_{1}F_{2}=0}$ and ${\small F}_{3}{\small F}_{4}={\small 0%
}$ then ${\small F}_{9}{\small =0}$ and ${\small F}_{10}{\small \neq 0}$ or $%
{\small F}_{9}{\small \neq 0}$ and ${\small F}_{10}{\small =0}$.

$\#1:$ If ${\small F_{1}F_{2}=0}$ and ${\small F}_{3}{\small F}_{4}={\small 0%
}$ then ${\small F}_{9}{\small =F}_{10}{\small =0}$.

$\#2:$ If ${\small F_{1}F}_{2}{\small =0}$ then ${\small F_{9}=F_{10}\neq 0}$%
.

$\#3:$ If ${\small F_{3}F}_{4}{\small =0}$ then ${\small F_{9}=F_{10}\neq 0}$%
.

$\#4:$ If $F_{i}=F_{j}=F_{k}=0$, where $1\leq i<j<k\leq 4$, then $%
|F_{9}|+|F_{10}|\neq 0$ and $F_{9}F_{10}=0.$

Table 2.2 The properties of $F_{i}$ for true entanglement classes

\begin{tabular}{ccc}
{\small classes} & ${\small F}_{i}$ & ${\small F}_{i}=0,{\small i=}$ \\ 
{\small \ }$|W\rangle $ &  & all $i$ \\ 
$|\chi _{4}\rangle $ & ${\small |F_{5}|+|F_{6}|\neq 0,|F_{7}|+|F_{8}|\neq 0,}%
\#0$ &  \\ 
$|\upsilon _{4}\rangle $ & ${\small |F}_{1}{\small |+|F}_{2}{\small |\neq
0,|F}_{3}{\small |+|F}_{4}{\small |\neq 0,|F}_{7}{\small |+|F}_{8}{\small %
|\neq 0,\#1}$ & ${\small i=5,6}$ \\ 
$|\varpi _{4}\rangle $ & ${\small |F}_{1}{\small |+|F}_{2}{\small |\neq
0,|F_{5}|+|F_{6}|\neq 0,|F_{7}|+|F_{8}|\neq 0,F}_{9}{\small =F}_{10},{\small %
F}_{1}{\small F}_{2}{\small =(F}_{9}{\small )}^{2}$ & ${\small 3,4}$ \\ 
{\small \ }$|\psi _{4}\rangle $ & ${\small F}_{9}{\small =F}_{10},{\small \#5%
}$ &  \\ 
{\small \ }$|\phi _{4}\rangle $ & $\#0$ &  \\ 
{\small \ }$|\mu _{4}\rangle $ & ${\small F}_{9}{\small =F}_{10}{\small ,\#5}
$ &  \\ 
$|\varphi _{4}\rangle $ & ${\small |F}_{1}{\small |+|F}_{2}{\small |\neq
0,|F_{5}|+|F_{6}|\neq 0,F}_{9}{\small =F}_{10}{\small ,F}_{1}{\small F}_{2}%
{\small =(F}_{9}{\small )}^{2}$ & ${\small 3,4,7,8}$ \\ 
$|\zeta _{4}\rangle $ & ${\small |F}_{1}{\small |+|F}_{2}{\small |\neq 0,|F}%
_{3}{\small |+|F}_{4}{\small |\neq 0,\#1}$ & ${\small 5,6,7,8}$ \\ 
$|\vartheta _{4}\rangle $ & ${\small |F}_{1}{\small |+|F}_{2}{\small |\neq
0,|F}_{7}{\small |+|F}_{8}{\small |\neq 0,F}_{9}{\small =F}_{10}{\small ,F}%
_{1}{\small F}_{2}{\small =(F}_{9}{\small )}^{2}$ & ${\small 3,4,5,6}$ \\ 
$|\tau _{4}\rangle $ & ${\small |F}_{3}{\small |+|F}_{4}{\small |\neq
0,|F_{5}|+|F_{6}|\neq 0,F}_{9}{\small =F}_{10}{\small ,F}_{3}{\small F}_{4}%
{\small =(F}_{9}{\small )}^{2}$ & ${\small 1,2,7,8}$ \\ 
$|\varrho _{4}\rangle $ & ${\small |F}_{3}{\small |+|F}_{4}{\small |\neq 0,|F%
}_{7}{\small |+|F}_{8}{\small |\neq 0,F}_{9}{\small =F}_{10}{\small ,F}_{3}%
{\small F}_{4}{\small =(F}_{9}{\small )}^{2}$ & ${\small 1,2,5,6}$ \\ 
$|\iota _{4}\rangle $ & ${\small |F}_{5}{\small |+|F}_{6}{\small |\neq 0,|F}%
_{7}{\small |+|F}_{8}{\small |\neq 0,}$ & ${\small 1,2,3,4,9,10}$ \\ 
{\small \ }$|\omega _{4}\rangle $ & ${\small |F}_{3}{\small |+|F}_{4}{\small %
|\neq 0,|F_{5}|+|F_{6}|\neq 0,|F_{7}|+|F_{8}|\neq 0,F}_{9}{\small =F}_{10}%
{\small ,F}_{3}{\small F}_{4}{\small =(F}_{9}{\small )}^{2}$ & ${\small i=1,2%
}$%
\end{tabular}

$\#5:$ If $F_{i}=F_{j}=F_{k}=0$, where $1\leq i<j<k\leq 4,$ then $F_{9}\neq
0 $.

Table 3.1. The properties of $D_{i}$ and $F_{i}$\ for the true entanglement
states

\begin{tabular}{ccccc}
{\small States} & $D_{1}$ & $D_{2}$ & $D_{3}$ & $F_{i}\neq 0,$ when $i=$ \\ 
$|GHZ\rangle $ & $=0$ & $=0$ & $=0$ & $9$ \\ 
$|C_{4}\rangle $ & $\neq 0$ & $\neq 0$ & $\neq 0$ & $9$ \\ 
$|\kappa _{4}\rangle $ & $=0$ & $=0$ & $=0$ & $9$ \\ 
$|E_{4}\rangle $ & $=0$ & $=0$ & $=0$ & $9$ \\ 
$|L_{4}\rangle $ & $=0$ & $=0$ & $=0$ & $9$ \\ 
$|H_{4}\rangle $ & $=0$ & $=0$ & $=0$ & $9$ \\ 
$|\lambda _{4}\rangle $ & $=0$ & $\neq 0$ & $=0$ & $10$ \\ 
$|M_{4}\rangle $ & $=0$ & $=0$ & $=0$ & $9$ \\ 
$|\pi _{4}\rangle $ & $\neq 0$ & $=0$ & $=0$ & $1,6,9,10,$ \\ 
$|\theta _{4}\rangle $ & $\neq 0$ & $=0$ & $=0$ & $4,7,9,10$ \\ 
$|\sigma _{4}\rangle $ & $=0$ & $\neq 0$ & $=0$ & $2,3$ \\ 
$|\rho _{4}\rangle $ & $=0$ & $\neq 0$ & $=0$ & $6,7$ \\ 
$|\xi _{4}\rangle $ & $=0$ & $=0$ & $\neq 0$ & $2,7,9,10,$ \\ 
$|\epsilon _{4}\rangle $ & $=0$ & $=0$ & $\neq 0$ & $3,6,9,10,$%
\end{tabular}

Table 3.2 The properties of $D_{i}$ and $F_{i}$\ for the true entanglement
states

\begin{tabular}{ccccc}
{\small States} & $D_{1}$ & $D_{2}$ & $D_{3}$ & $F_{i}\neq 0,$ when $i=$ \\ 
$|W\rangle $ & $=0$ & $=0$ & $=0$ &  \\ 
$|\chi _{4}\rangle $ & $=0$ & $\neq 0$ & $=0$ & $6,7,9$ \\ 
$|\upsilon _{4}\rangle $ & $=0$ & $=0$ & $=0$ & $1,3,8$ \\ 
$|\varpi _{4}\rangle $ & $=0$ & $=0$ & $=0$ & $1,5,7$ \\ 
$|\psi _{4}\rangle $ & $\neq 0$ & $=0$ & $=0$ & $3,4,9,10$ \\ 
$|\phi _{4}\rangle $ & $=0$ & $\neq 0$ & $=0$ & $9$ \\ 
$|\mu _{4}\rangle $ & $=0$ & $=0$ & $\neq 0$ & $9,10$ \\ 
$|\varphi _{4}\rangle $ & $=0$ & $=0$ & $=0$ & $1,6$ \\ 
$|\zeta _{4}\rangle $ & $=0$ & $=0$ & $=0$ & $2,3$ \\ 
$|\vartheta _{4}\rangle $ & $=0$ & $=0$ & $=0$ & $1,8$ \\ 
$|\tau _{4}\rangle $ & $=0$ & $=0$ & $=0$ & $3,6$ \\ 
$|\varrho _{4}\rangle $ & $=0$ & $=0$ & $=0$ & $3,8$ \\ 
$|\iota _{4}\rangle $ & $=0$ & $=0$ & $=0$ & $5,8$ \\ 
$|\omega _{4}\rangle $ & $=0$ & $=0$ & $=0$ & $4,5,7,$%
\end{tabular}

Table 4. The invariants of the degenerated entanglement classes

\begin{tabular}{cccccc}
Classes & $IV$ & $D_{1}$ & $D_{2}$ & $D_{3}$ & $F$ \\ 
$|GHZ\rangle _{123}\otimes (s|0\rangle +t|1\rangle )_{4}$ & $=0$ & $=0$ & $%
=0 $ & $=0$ & $>0$ \\ 
$|GHZ\rangle _{124}\otimes (s|0\rangle +t|1\rangle )_{3}$ & $=0$ & $=0$ & $%
=0 $ & $=0$ & $>0$ \\ 
$|GHZ\rangle _{134}\otimes (s|0\rangle +t|1\rangle )_{2}$ & $=0$ & $=0$ & $%
=0 $ & $=0$ & $>0$ \\ 
$(s|0\rangle +t|1\rangle )_{1}\otimes |GHZ\rangle _{234}$ & $=0$ & $=0$ & $%
=0 $ & $=0$ & $>0$ \\ 
$|W\rangle \otimes (s|0\rangle +t|1\rangle )_{i}$ & $=0$ & $=0$ & $=0$ & $=0$
& $=0$ \\ 
$|GHZ\rangle _{12}\otimes |GHZ\rangle _{34}$ & $\neq 0$ & $=0$ & opt & $=0$
& $=0$ \\ 
$|GHZ\rangle _{13}\otimes |GHZ\rangle _{24}$ & $\neq 0$ & opt & $=0$ & $=0$
& $>0$ \\ 
$|GHZ\rangle _{14}\otimes |GHZ\rangle _{23}$ & $\neq 0$ & $=0$ & $=0$ & opt
& $>0$ \\ 
{\small only two qubits are entangled} & $=0$ & $=0$ & $=0$ & $=0$ & $=0$ \\ 
separate states & $=0$ & $=0$ & $=0$ & $=0$ & $=0$%
\end{tabular}

In $|W\rangle \otimes (s|0\rangle +t|1\rangle )_{i}$, $i=1,2,3,4.$

Table 5. The properties of $F_{i}$ for the degenerated entanglement classes

\begin{tabular}{ccc}
Classes & $F_{i}=0$ &  \\ 
$|GHZ\rangle _{123}\otimes (s|0\rangle +t|1\rangle )_{4}$ & $i\neq 7,8$ & $%
|F_{7}|$ $+$ $|F_{8}|\neq 0$ \\ 
$|GHZ\rangle _{124}\otimes (s|0\rangle +t|1\rangle )_{3}$ & $i\neq 5,6$ & $%
|F_{5}|+|F_{6}|\neq 0$ \\ 
$|GHZ\rangle _{134}\otimes (s|0\rangle +t|1\rangle )_{2}$ & $i\neq 3,4,9,10$
& $|F_{3}|$ $+$ $|F_{4}|\neq 0,F_{9}=F_{10},F_{3}F_{4}=(F_{9})^{2}$ \\ 
$(s|0\rangle +t|1\rangle )_{1}\otimes |GHZ\rangle _{234}$ & $i\neq 1,2,9,10$
& $|F_{1}|$ $+$ $|F_{2}|\neq 0,F_{9}=F_{10},F_{1}F_{2}=(F_{9})^{2}$ \\ 
$|W\rangle \otimes (s|0\rangle +t|1\rangle )_{i}$ & All $F_{i}=0$ &  \\ 
$|GHZ\rangle _{12}\otimes |GHZ\rangle _{34}$ & All $F_{i}=0$ &  \\ 
$|GHZ\rangle _{13}\otimes |GHZ\rangle _{24}$ & $i\neq 9,10$ & $%
F_{9}=F_{10}\neq 0$ \\ 
$|GHZ\rangle _{14}\otimes |GHZ\rangle _{23}$ & $i\neq 9,10$ & $%
F_{9}=F_{10}\neq 0$ \\ 
{\small only two qubits are entangled} & All $F_{i}=0$ &  \\ 
separate states & All $F_{i}=0$ & 
\end{tabular}

Table 6. The properties of $D_{i}$ for two $GHZ$ pairs

\begin{tabular}{cccccccc}
& states &  & $D_{1}$ & $D_{2}$ & $D_{3}$ &  &  \\ 
& $|GHZ\rangle _{12}\otimes |GHZ\rangle _{34}$ &  & $=0$ & $\neq 0$ & $=0$ & 
&  \\ 
& $|GHZ\rangle _{13}\otimes |GHZ\rangle _{24}$ &  & $\neq 0$ & $=0$ & $=0$ & 
&  \\ 
& $|GHZ\rangle _{14}\otimes |GHZ\rangle _{23}$ &  & $=0$ & $=0$ & $\neq 0$ & 
& 
\end{tabular}

\section{Semi-invariants of $n$-qubits}

Definition:

The semi-invariants of a pure state of $n$-qubits is 
\begin{eqnarray}
F &=&4(\sum_{odd(i+j)}\left\vert
(a_{i}a_{j}+a_{k}a_{l}-a_{p}a_{q}-a_{r}a_{s})^{2}-4(a_{i}a_{j-1}-a_{p}a_{q-1})(a_{k}a_{l+1}-a_{r}a_{s+1})\right\vert
\nonumber \\
&&+\sum_{Even(i+j)}\left\vert
(a_{i}a_{j}+a_{k}a_{l}-a_{p}a_{q}-a_{r}a_{s})^{2}-4(a_{i}a_{j-2}-a_{p}a_{q-2})(a_{k}a_{l+2}-a_{r}a_{s+2})\right\vert ),
\nonumber \\
&&
\end{eqnarray}

where 
\begin{eqnarray}
&&i<j,k<l,p<q, r<s, i<k<p<r  \nonumber \\
&&i+j=k+l=p+q=r+s, i\oplus j=k\oplus l=p\oplus q=r\oplus s.  \label{concur}
\end{eqnarray}

For example, $F\ $contains the terms in which $i+j=7,11,13,15,17,19$ and $23$
and the terms in which $i+j=14$ and $16,$ but $F$ does not contain the terms
in which $i+j=8,9,10,12,18,20,21$ or $22$.

Remark:

We can consider that Eq. (\ref{concur}) is a special partition of an integer.

Lemma 1.

Let $(|0\rangle +|2^{n}-1\rangle )/2$ be state $|GHZ\rangle $ of $n$-qubits.
Then $F$ of state $|GHZ\rangle $ \ does not vanish.

Proof. This is because the following term does not vanish.

$\left|
((a_{0}a_{2^{n}-1}-a_{2}a_{2^{n}-3})+(a_{1}a_{2^{n}-2}-a_{3}a_{2^{n}-4}))^{2}-4(a_{0}a_{2^{n}-2}-a_{2}a_{2^{n}-4})(a_{1}a_{2^{n}-1}-a_{3}a_{2^{n}-3})\right| 
$

$=\left| a_{0}a_{2^{n}-1}\right| =1/4.$

Notice that other terms vanish.

Lemma 2.

Let $|W\rangle =(|0...01\rangle +|0...010\rangle +|0...0100\rangle +...)/%
\sqrt{n}$, where the amplitudes $a_{2^{j}}=1/\sqrt{n}$, where $%
j=0,1,...,(n-1)$, and other amplitudes $a_{i}=0$. Then $F$ of $|W\rangle $
vanishes.

Proof.

(1). We show that $a_{i}a_{j}=a_{k}a_{l}=a_{p}a_{q}=a_{r}a_{s}=0$. If $%
a_{i}a_{j}\neq 0$, then $i=2^{m}$ and $j=2^{n}$, where $m<n$. Clearly, we
can not find $k$ or $l$ such that $k=2^{s}$ and $l=2^{t}$ and $2^{m}\oplus
2^{n}=2^{s}\oplus 2^{t}$.

(2). We show that $(a_{i}a_{j-1}-a_{p}a_{q-1})(a_{k}a_{l+1}-a_{r}a_{s+1})=0$%
. It is enough to illustrate $%
a_{i}a_{j-1}=a_{p}a_{q-1}=a_{k}a_{l+1}=a_{r}a_{s+1}=0$. Assume that $i=2^{m}$
and $j-1=2^{n}$ and $k=2^{s}$ and $l+1=2^{t}$. Since $i+j=k+l$, $%
2^{m}+2^{n}+2=2^{s}+2^{t}$. Since $i\oplus j=k\oplus l$, then\ $i\oplus
(j-1)=k\oplus (l+1)$, i.e., $2^{m}\oplus 2^{n}=2^{s}\oplus 2^{t}$. It is not
possible for $2^{m}+2^{n}+2=2^{s}+2^{t}$ and $2^{m}\oplus 2^{n}=2^{s}\oplus
2 $ to both hold.

(3). As well, we can show that $%
(a_{i}a_{j-2}-a_{p}a_{q-2})(a_{k}a_{l+2}-a_{r}a_{s+2})=0.$

Conclusively, $F$ vanishes.

\textbf{Summary}

In this paper, we define the SLOCC invariant and semi-invariants for
four-qubits. By means of the invariant and semi-invariants we can determine
if two states are inequivalent. Then, we show that there are infinite $SL$%
-classes and at least 28 distinct true SLOCC entanglement classes. It seems
that there should be more true SLOCC entanglement classes. The invariant and
semi-invariants only require simple arithmetic operations. The ideas can be
extended to five or more-qubits for SLOCC classification. In this paper, we
also give the exact recursive formulas of the number of the degenerated
SLOCC classes of $n$-qubits. By the recursive formula, for six-qubits, there
are $6\ast t(5)+30\ast t(4)+276$ distinct degenerated SLOCC entanglement
classes, where $t(5)$ is the number of the true SLOCC entanglement classes
for five-qubits.

\textbf{Appendix A}

Let $P=\det^{2}(\beta )\det^{2}(\delta )\det^{2}(\gamma )$, $%
Q=\det^{2}(\alpha )\det^{2}(\gamma )\det^{2}(\delta )$, $R=\det^{2}(\alpha
)\det^{2}(\beta )\det^{2}(\delta )$, $S=\det^{2}(\alpha )\det^{2}(\gamma
)\det^{2}(\beta )$, $T=\det (\alpha )\det (\beta )\det (\gamma )\det (\delta
)$.

We will list $G$, $D_{i}$ and $F_{i}$ which are not zero as follows.

1. Class $|GHZ\rangle $

(1). If $F_{i}=0$ then $|F_{9}|+|F_{10}|\neq 0$, $i=1,2,3$ and $4$.

\noindent (2). $IV=-1/2\ast T$.

Proof of (1):

Let $|\psi ^{\prime }\rangle $ in Eq. (\ref{eq1}) be state $|GHZ\rangle $.
By solving matrix equation in Eq. (\ref{eq1}), we obtain \ the amplitudes $%
a_{i}$. By computing, we obtain the following $F_{i}$.

$F_{1}=\allowbreak \frac{1}{4}\alpha _{1}^{2}\alpha _{2}^{2}P$, $F_{2}=\frac{%
1}{4}\alpha _{3}^{2}\alpha _{4}^{2}P$, $F_{3}=\frac{1}{4}\beta _{1}^{2}\beta
_{2}^{2}Q$, $F_{4}=\allowbreak \frac{1}{4}\beta _{3}^{2}\beta _{4}^{2}Q$, $%
F_{5}=\frac{1}{4}\gamma _{2}^{2}\gamma _{1}^{2}R$, $F_{6}=\allowbreak \frac{1%
}{4}\gamma _{4}^{2}\gamma _{3}^{2}R$, $F_{7}=\allowbreak \frac{1}{4}\delta
_{1}^{2}\delta _{2}^{2}S$, $F_{8}=\frac{1}{4}\delta _{3}^{2}\delta _{4}^{2}S$%
,

$F_{9}=\allowbreak \frac{1}{4}\left( \alpha _{1}\beta _{1}\alpha _{4}\beta
_{4}-\beta _{3}\alpha _{3}\beta _{2}\alpha _{2}\right) ^{2}\det^{2}(\delta
)\det^{2}(\gamma )$,

$F_{10}=\frac{1}{4}\left( -\alpha _{1}\beta _{3}\alpha _{4}\beta _{2}+\beta
_{1}\alpha _{3}\beta _{4}\alpha _{2}\right) ^{2}\allowbreak \det^{2}(\delta
)\det^{2}(\gamma )$.

Assume $F_{1}=0$ and $F_{3}=0$. Then $\alpha _{1}\alpha _{2}=0$. Without
loss of generality, let us consider $\alpha _{1}=0$. This implies $\alpha
_{2}\alpha _{3}\neq 0$ since $\alpha $ is invertible. Thus, $%
F_{9}=\allowbreak \frac{1}{4}\left( \beta _{2}\beta _{3}\alpha _{2}\alpha
_{3}\right) ^{2}\det^{2}(\delta )\det^{2}(\gamma )$ and $F_{10}=\frac{1}{4}%
\left( \beta _{1}\beta _{4}\alpha _{2}\alpha _{3}\right) ^{2}\allowbreak
\det^{2}(\delta )\det^{2}(\gamma )$. If $\beta _{1}=0$ then $F_{10}=0$ and $%
F_{9}\neq 0$ because $\beta $ is invertible. If $\beta _{2}=0$, then $%
F_{9}=0 $ and $F_{10}\neq 0$. Similarly, it is easy to verify other cases.

2. Class $|C_{4}\rangle $

(1). If $F_{i}=F_{j}=F_{k}=0$, where $1\leq i<j<k\leq 4$, then $%
|F_{9}|+|F_{10}|\neq 0$ and $F_{9}F_{10}=0.$

\noindent (2). $IV=-1/2\ast T$.

$D_{1}=(-1/36)((\alpha _{2}\alpha _{3}+\alpha _{1}\alpha _{4})(\gamma
_{2}\gamma _{3}+\gamma _{1}\gamma _{4})+\alpha _{2}\alpha _{4}\gamma
_{1}\gamma _{3}+\alpha _{1}\alpha _{3}\gamma _{2}\gamma _{4})\det (\alpha
)\det^{2}(\beta )\det (\gamma )\det^{2}(\delta )$,

$D_{2}=(1/36)((\alpha _{2}\alpha _{3}+\alpha _{1}\alpha _{4})(\beta
_{2}\beta _{3}+\beta _{1}\beta _{4})+\alpha _{2}\alpha _{4}\beta _{1}\beta
_{3}+\alpha _{1}\alpha _{3}\beta _{2}\beta _{4})\det (\alpha )\det (\beta
)\det^{2}(\gamma )\det^{2}(\delta )$,

$D_{3}=(1/36)((\alpha _{2}\alpha _{3}+\alpha _{1}\alpha _{4})(\delta
_{2}\delta _{3}+\delta _{1}\delta _{4})+\alpha _{2}\alpha _{4}\delta
_{1}\delta _{3}+\alpha _{1}\alpha _{3}\delta _{2}\delta _{4})\det (\alpha
)\det^{2}(\beta )\det^{2}(\gamma )\det (\delta )$.

Proof of (1):

Let $|\psi ^{\prime }\rangle $ in Eq. (\ref{eq1}) be state $|C_{4}\rangle $.
By solving matrix equation in Eq. (\ref{eq1}), we obtain the amplitudes $%
a_{i}$. Then, we obtain the following $F_{i}$.

$F_{1}$ to $F_{8}$ can be obtained from the ones of class $|GHZ\rangle $
above by replacing $1/4$ by $(-1/12)$, respectively.

$F_{9}=(1/36)(-4\alpha _{2}^{2}\alpha _{3}\alpha _{4}\beta _{1}\beta
_{2}\beta _{3}^{2}+4\alpha _{1}\alpha _{2}\alpha _{4}^{2}\beta _{1}\beta
_{2}\beta _{3}^{2}+\alpha _{2}^{2}\alpha _{3}^{2}\beta _{2}^{2}\beta
_{3}^{2}-4\alpha _{1}\alpha _{2}\alpha _{3}\alpha _{4}\beta _{2}^{2}\beta
_{3}^{2}+4\alpha _{2}^{2}\alpha _{3}\alpha _{4}\beta _{1}^{2}\beta _{3}\beta
_{4}-4\alpha _{1}\alpha _{2}\alpha _{4}^{2}\beta _{1}^{2}\beta _{3}\beta
_{4}-4\alpha _{2}^{2}\alpha _{3}^{2}\beta _{1}\beta _{2}\beta _{3}\beta
_{4}+14\alpha _{1}\alpha _{2}\alpha _{3}\alpha _{4}\beta _{1}\beta _{2}\beta
_{3}\beta _{4}-4\alpha _{1}^{2}\alpha _{4}^{2}\beta _{1}\beta _{2}\beta
_{3}\beta _{4}-4\alpha _{1}\alpha _{2}\alpha _{3}^{2}\beta _{2}^{2}\beta
_{3}\beta _{4}+4\alpha _{1}^{2}\alpha _{3}\alpha _{4}\beta _{2}^{2}\beta
_{3}\beta _{4}-4\alpha _{1}\alpha _{2}\alpha _{3}\alpha _{4}\beta
_{1}^{2}\beta _{4}^{2}+\alpha _{1}^{2}\alpha _{4}^{2}\beta _{1}^{2}\beta
_{4}^{2}+4\alpha _{1}\alpha _{2}\alpha _{3}^{2}\beta _{1}\beta _{2}\beta
_{4}^{2}-4\alpha _{1}^{2}\alpha _{3}\alpha _{4}\beta _{1}\beta _{2}\beta
_{4}^{2})\det^{2}(\delta )\det^{2}(\gamma )$,

$F_{10}=(1/36)(4\alpha _{2}^{2}\alpha _{3}\alpha _{4}\beta _{1}\beta
_{2}\beta _{3}^{2}-4\alpha _{1}\alpha _{2}\alpha _{4}^{2}\beta _{1}\beta
_{2}\beta _{3}^{2}-4\alpha _{1}\alpha _{2}\alpha _{3}\alpha _{4}\beta
_{2}^{2}\beta _{3}^{2}+\alpha _{1}^{2}\alpha _{4}^{2}\beta _{2}^{2}\beta
_{3}^{2}-4\alpha _{2}^{2}\alpha _{3}\alpha _{4}\beta _{1}^{2}\beta _{3}\beta
_{4}+4\alpha _{1}\alpha _{2}\alpha _{4}^{2}\beta _{1}^{2}\beta _{3}\beta
_{4}-4\alpha _{2}^{2}\alpha _{3}^{2}\beta _{1}\beta _{2}\beta _{3}\beta
_{4}+14\alpha _{1}\alpha _{2}\alpha _{3}\alpha _{4}\beta _{1}\beta _{2}\beta
_{3}\beta _{4}-4\alpha _{1}^{2}\alpha _{4}^{2}\beta _{1}\beta _{2}\beta
_{3}\beta _{4}+4\alpha _{1}\alpha _{2}\alpha _{3}^{2}\beta _{2}^{2}\beta
_{3}\beta _{4}-4\alpha _{1}^{2}\alpha _{3}\alpha _{4}\beta _{2}^{2}\beta
_{3}\beta _{4}+\alpha _{2}^{2}\alpha _{3}^{2}\beta _{1}^{2}\beta
_{4}^{2}-4\alpha _{1}\alpha _{2}\alpha _{3}\alpha _{4}\beta _{1}^{2}\beta
_{4}^{2}-4\alpha _{1}\alpha _{2}\alpha _{3}^{2}\beta _{1}\beta _{2}\beta
_{4}^{2}+4\alpha _{1}^{2}\alpha _{3}\alpha _{4}\beta _{1}\beta _{2}\beta
_{4}^{2})\det^{2}(\delta )\det^{2}(\gamma )$.

Let us prove that if $F_{1}=F_{3}=F_{4}=0$ then $|F_{9}|+|F_{10}|\neq 0$ and 
$F_{9}F_{10}=0$. The proofs for other cases are similar.

Assume that $F_{1}=0$. Then there are two cases: case 1, $\alpha _{1}=0$ and
case 2, $\alpha _{2}=0$.

Case 1. $\alpha _{1}=0$. In this case, $\alpha _{2}\alpha _{3}\neq 0$. Thus,

$F_{9}=(1/36)\alpha _{2}^{2}\alpha _{3}\beta _{3}\left( \alpha _{3}\beta
_{2}^{2}\beta _{3}+4\beta _{1}^{2}\alpha _{4}\beta _{4}-4\beta _{1}\alpha
_{3}\beta _{2}\beta _{4}-4\beta _{1}\beta _{2}\alpha _{4}\beta _{3}\right)
\allowbreak \det^{2}(\delta )\det^{2}(\gamma )$,

$F_{10}=(1/36)\alpha _{2}^{2}\beta _{1}\alpha _{3}\left( \beta _{1}\alpha
_{3}\beta _{4}^{2}+4\beta _{2}\alpha _{4}\beta _{3}^{2}-4\beta _{1}\alpha
_{4}\beta _{3}\beta _{4}-4\alpha _{3}\beta _{2}\beta _{3}\beta _{4}\right)
\allowbreak \det^{2}(\delta )\det^{2}(\gamma )$.

Since $F_{3}=F_{4}=0$, there are two cases.

Case 1.1. $\beta _{1}=\beta _{4}=0$. Then $\beta _{2}\beta _{3}\neq 0$.

$F_{9}=(1/36)\allowbreak \alpha _{2}^{2}\alpha _{3}^{2}\beta _{2}^{2}\beta
_{3}^{2}\allowbreak \det^{2}(\delta )\det^{2}(\gamma )\neq 0$,

$F_{10}=0$.

Case 1.2. $\beta _{2}=\beta _{3}=0$. Then $\beta _{1}\beta _{4}\neq 0$.

$F_{9}=0$,

$F_{10}=(1/36)\alpha _{2}^{2}\alpha _{3}^{2}\left( \beta _{1}^{2}\beta
_{4}^{2}\right) \allowbreak \det^{2}(\delta )\det^{2}(\gamma )\neq 0$.

Case 2. $\alpha _{2}=0$. In this case, $\alpha _{1}\alpha _{4}\neq 0$. Thus,

$F_{9}=(1/36)\allowbreak \alpha _{1}^{2}\alpha _{4}\beta _{4}\left( 4\alpha
_{3}\beta _{2}^{2}\beta _{3}+\beta _{1}^{2}\alpha _{4}\beta _{4}-4\beta
_{1}\alpha _{3}\beta _{2}\beta _{4}-4\beta _{1}\beta _{2}\alpha _{4}\beta
_{3}\right) \allowbreak \det^{2}(\delta )\det^{2}(\gamma )$,

$F_{10}=(1/36)\allowbreak \alpha _{1}^{2}\beta _{2}\alpha _{4}\left( 4\beta
_{1}\alpha _{3}\beta _{4}^{2}+\beta _{2}\alpha _{4}\beta _{3}^{2}-4\beta
_{1}\alpha _{4}\beta _{3}\beta _{4}-4\alpha _{3}\beta _{2}\beta _{3}\beta
_{4}\right) \allowbreak \det^{2}(\delta )\det^{2}(\gamma )$.

Since $F_{3}=F_{4}=0$, there are two cases.

Case 1.1. $\beta _{1}=\beta _{4}=0$. Then $\beta _{2}\beta _{3}\neq 0$.

$F_{9}=0$,

$F_{10}=(1/36)\allowbreak \alpha _{1}^{2}\alpha _{4}^{2}\beta _{2}^{2}\beta
_{3}^{2}\allowbreak \det^{2}(\delta )\det^{2}(\gamma )\neq 0$.

Case 1.2. $\beta _{2}=\beta _{3}=0$. Then $\beta _{1}\beta _{4}\neq 0$.

$F_{9}=(1/36)\allowbreak \alpha _{1}^{2}\alpha _{4}^{2}\left( \beta
_{1}^{2}\beta _{4}^{2}\right) \allowbreak \det^{2}(\delta )\det^{2}(\gamma
)\neq 0$,

$F_{10}=0$.

3. Class $|\kappa _{4}\rangle $

(1). $IV=1/4\ast T$.

$\ D_{1}=(1/16)\alpha _{2}\alpha _{4}\gamma _{2}\gamma _{4}\det (\alpha
)\det^{2}(\beta )\det (\gamma )\det^{2}(\delta )$,$\ $

$D_{2}=(1/16)\alpha _{1}\alpha _{3}\beta _{1}\beta _{3}\det (\alpha )\det
(\beta )\det^{2}(\gamma )\det^{2}(\delta )$.

$\ $(2).$\#0$

Proof of (2):

$F_{i}$ can be obtained from the $F_{i}$ of class $|GHZ\rangle $ by
replacing $1/4$ by $1/16$. So the proof is similar to the one of class $%
|GHZ\rangle $.

4. Class $|E_{4}\rangle $

(1).$\ IV=1/4\ast T.$

$\ D_{1}=-(1/16)\alpha _{1}\alpha _{3}\gamma _{1}\gamma _{3}\det (\alpha
)\det^{2}(\beta )\det (\gamma )\det^{2}(\delta )$,$\ $

$D_{3}=-(1/16)\alpha _{2}\alpha _{4}\delta _{2}\delta _{4}\det (\alpha
)\det^{2}(\beta )\det^{2}(\gamma )\det (\delta )$.

(2).$\ \#0$

Proof of (2):

$F_{i}$ are the same as the $F_{i}$ of class $|\kappa _{4}\rangle $. So the
proof is similar to the one of $|\kappa _{4}\rangle $.

5. Class $|L_{4}\rangle $

(1). $IV=1/4\ast T.$

$\ \ D_{2}=(1/16)\alpha _{1}\alpha _{3}\beta _{1}\beta _{3}\det (\alpha
)\det (\beta )\det^{2}(\gamma )\det^{2}(\delta )$,$\ $

$\ D_{3}=-(1/16)\alpha _{2}\alpha _{4}\delta _{2}\delta _{4}\det (\alpha
)\det^{2}(\beta )\det^{2}(\gamma )\det (\delta )$.

(2). $\#0$

Proof of (2):

$\ F_{i}$ are the same as the $F_{i}$ of class $|\kappa _{4}\rangle $.

6. Class $|H_{4}\rangle $

(1). $IV=-1/3\ast T$. $D_{1}=-(1/9)\alpha _{1}\alpha _{3}\gamma _{2}\gamma
_{4}\det (\alpha )\det^{2}(\beta )\det (\gamma )\det^{2}(\delta )$.

(2). $\#0$

Proof of (2):

$F_{i}$ can be obtained from the $F_{i}$ of class $|GHZ\rangle $ by
replacing $1/4$ by $1/9$.

7. Class $|\lambda _{4}\rangle $

(1). $IV=-1/3\ast T$. $D_{2}=(1/9)\alpha _{1}\alpha _{3}\beta _{2}\beta
_{4}\det (\alpha )\det (\beta )\det^{2}(\gamma )\det^{2}(\delta )$.

(2). $\#0$

Proof of (2):

\noindent\ $\ \ F_{1}$ to $F_{8}$ can be obtained from $F_{1}$ to $F_{8}$ of
class $|GHZ\rangle $ by replacing $1/4$ by $1/9$, respectively.

$F_{9}=(1/9)\allowbreak \left( \alpha _{1}\beta _{2}\alpha _{4}\beta
_{3}-\beta _{1}\alpha _{2}\beta _{4}\alpha _{3}\right) ^{2}\det^{2}(\gamma
)\det^{2}(\delta ),$

$F_{10}=(1/9)\left( -\alpha _{1}\beta _{1}\alpha _{4}\beta _{4}+\beta
_{2}\alpha _{3}\beta _{3}\alpha _{2}\right) ^{2}\det^{2}(\gamma
)\det^{2}(\delta )\allowbreak .$ \ 

The next argument is the same as the one for class $|GHZ\rangle $.

8. Class $M_{4}$

(1). $IV=-1/3\ast T$. $D_{3}=(1/9)\alpha _{1}\alpha _{3}\delta _{2}\delta
_{4}\det (\alpha )\det^{2}(\beta )\det^{2}(\gamma )\det (\delta )$.

(2). $\#0$

Proof of (2):

$F_{i}$ are the same as the ones of class $|H_{4}\rangle $.

9. Class $|\pi _{4}\rangle $

(1). $|F_{1}|+|F_{2}|\neq 0,|F_{5}|+|F_{6}|\neq 0$, $\#2$.

\noindent\ \ \ (2). $IV=-1/3\ast T$.

$D_{1}=\frac{1}{36}(2\alpha _{1}\alpha _{3}\gamma _{1}\gamma _{3}+\alpha
_{2}\alpha _{3}\gamma _{2}\gamma _{3}+\alpha _{1}\alpha _{4}\gamma
_{2}\gamma _{3}+\alpha _{2}\alpha _{3}\gamma _{1}\gamma _{4}+\alpha
_{1}\alpha _{4}\gamma _{1}\gamma _{4}+2\alpha _{1}\alpha _{3}\gamma
_{2}\gamma _{4}+2\alpha _{2}\alpha _{4}\gamma _{2}\gamma _{4})\det (\alpha
)\det^{2}(\beta )\det (\gamma )\det^{2}(\delta )$

Proof of (1):

Let $|\psi ^{\prime }\rangle $ in Eq. (\ref{eq1})\ be state $|\pi
_{4}\rangle $. By solving matrix equation in Eq. (\ref{eq1}), we obtain the
amplitudes $a_{i}$. Then, we obtain the following $F_{i}$.

$F_{1}=\alpha _{1}^{4}P/9$, $F_{2}=\alpha _{3}^{4}P/9$, $F_{5}=\gamma
_{2}^{4}R/9$, $F_{6}=\gamma _{4}^{4}R/9$,

$F_{9}=\det (\beta )(-4\alpha _{1}\alpha _{2}\alpha _{3}^{2}\beta _{1}\beta
_{3}+4\alpha _{1}^{2}\alpha _{3}\alpha _{4}\beta _{1}\beta _{3}-4\alpha
_{1}^{2}\alpha _{3}^{2}\beta _{2}\beta _{3}-\alpha _{2}^{2}\alpha
_{3}^{2}\beta _{2}\beta _{3}+2\alpha _{1}\alpha _{2}\alpha _{3}\alpha
_{4}\beta _{2}\beta _{3}-\alpha _{1}^{2}\alpha _{4}^{2}\beta _{2}\beta
_{3}+4\alpha _{1}^{2}\alpha _{3}^{2}\beta _{1}\beta _{4}+\alpha
_{2}^{2}\alpha _{3}^{2}\beta _{1}\beta _{4}-2\alpha _{1}\alpha _{2}\alpha
_{3}\alpha _{4}\beta _{1}\beta _{4}+\alpha _{1}^{2}\alpha _{4}^{2}\beta
_{1}\beta _{4}+4\alpha _{1}\alpha _{2}\alpha _{3}^{2}\beta _{2}\beta
_{4}-4\alpha _{1}^{2}\alpha _{3}\alpha _{4}\beta _{2}\beta
_{4})\det^{2}(\gamma )\det^{2}(\delta )/36$,

$F_{10}=\det (\beta )(4\alpha _{1}\alpha _{2}\alpha _{3}^{2}\beta _{1}\beta
_{3}-4\alpha _{1}^{2}\alpha _{3}\alpha _{4}\beta _{1}\beta _{3}-4\alpha
_{1}^{2}\alpha _{3}^{2}\beta _{2}\beta _{3}-\alpha _{2}^{2}\alpha
_{3}^{2}\beta _{2}\beta _{3}+2\alpha _{1}\alpha _{2}\alpha _{3}\alpha
_{4}\beta _{2}\beta _{3}-\alpha _{1}^{2}\alpha _{4}^{2}\beta _{2}\beta
_{3}+4\alpha _{1}^{2}\alpha _{3}^{2}\beta _{1}\beta _{4}+\alpha
_{2}^{2}\alpha _{3}^{2}\beta _{1}\beta _{4}-2\alpha _{1}\alpha _{2}\alpha
_{3}\alpha _{4}\beta _{1}\beta _{4}+\alpha _{1}^{2}\alpha _{4}^{2}\beta
_{1}\beta _{4}-4\alpha _{1}\alpha _{2}\alpha _{3}^{2}\beta _{2}\beta
_{4}+4\alpha _{1}^{2}\alpha _{3}\alpha _{4}\beta _{2}\beta
_{4})\det^{2}(\gamma )\det^{2}(\delta )/36$.

Clearly, if $F_{1}=0$, then $\alpha _{1}=0$. Since $\alpha $ is invertible,
that $\alpha _{1}=0$ implies $\alpha _{3}\neq 0$. Thus, $F_{2}\neq 0$.
Therefore $|F_{1}|+|F_{2}|\neq 0$. As well, $|F_{5}|+|F_{6}|\neq 0$.

Next we prove if $F_{1}F_{2}=0$ then $F_{9}=F_{10}\neq 0$. Assume $F_{1}=0$.
Then $\alpha _{1}=0$. By substituting $\alpha _{1}=0$ into $F_{9}$ and $%
F_{10}$, $F_{9}=$ $F_{10}=$ $\alpha _{2}^{2}\alpha _{3}^{2}(\beta _{1}\beta
_{4}-\beta _{2}\beta _{3})/36$. Since $\alpha $ and $\beta $\ are
invertible, $F_{9}=$ $F_{10}\neq 0$. Similarly, we can argue $%
F_{9}=F_{10}\neq 0$ if $F_{2}=0$.

10. Class $|\theta _{4}\rangle $

(1). $IV=-1/3\ast T$.

$D_{1}=\frac{1}{36}(2\alpha _{1}\alpha _{3}\gamma _{1}\gamma _{3}+\alpha
_{2}\alpha _{3}\gamma _{2}\gamma _{3}+\alpha _{1}\alpha _{4}\gamma
_{2}\gamma _{3}+\alpha _{2}\alpha _{3}\gamma _{1}\gamma _{4}+\alpha
_{1}\alpha _{4}\gamma _{1}\gamma _{4}+2\alpha _{2}\alpha _{4}\gamma
_{2}\gamma _{4})\det (\alpha )\det^{2}(\beta )\det (\gamma )\det^{2}(\delta
) $.

\bigskip (2). $|F_{3}|+|F_{4}|\neq 0,|F_{7}|+|F_{8}|\neq 0,\#3$.

Proof.

Let $|\psi ^{\prime }\rangle $ in Eq. (\ref{eq1})\ be state $|\theta
_{4}\rangle $. By solving matrix equation in Eq. (\ref{eq1}), we obtain the
amplitudes $a_{i}$. Then, we obtain the following $F_{i}$.

$F_{3}=\frac{1}{9}\beta _{2}^{4}Q$, $F_{4}=\frac{1}{9}\beta _{4}^{4}Q$, $%
F_{7}=\frac{1}{9}\delta _{1}^{4}S$, $F_{8}=\frac{1}{9}\delta _{3}^{4}S$,

$F_{9}$ $=\frac{1}{36}(-\alpha _{2}\alpha _{3}\beta _{2}^{2}\beta
_{3}^{2}+\alpha _{1}\alpha _{4}\beta _{2}^{2}\beta _{3}^{2}+2\alpha
_{2}\alpha _{3}\beta _{1}\beta _{2}\beta _{3}\beta _{4}-2\alpha _{1}\alpha
_{4}\beta _{1}\beta _{2}\beta _{3}\beta _{4}+4\alpha _{1}\alpha _{3}\beta
_{2}^{2}\beta _{3}\beta _{4}-4\alpha _{2}\alpha _{4}\beta _{2}^{2}\beta
_{3}\beta _{4}-\alpha _{2}\alpha _{3}\beta _{1}^{2}\beta _{4}^{2}+\alpha
_{1}\alpha _{4}\beta _{1}^{2}\beta _{4}^{2}-4\alpha _{1}\alpha _{3}\beta
_{1}\beta _{2}\beta _{4}^{2}+4\alpha _{2}\alpha _{4}\beta _{1}\beta
_{2}\beta _{4}^{2}-4\alpha _{2}\alpha _{3}\beta _{2}^{2}\beta
_{4}^{2}+4\alpha _{1}\alpha _{4}\beta _{2}^{2}\beta _{4}^{2})\det (\alpha
)\det^{2}(\gamma )\det^{2}(\delta )$,

$F_{10}=\frac{1}{36}(-\alpha _{2}\alpha _{3}\beta _{2}^{2}\beta
_{3}^{2}+\alpha _{1}\alpha _{4}\beta _{2}^{2}\beta _{3}^{2}+2\alpha
_{2}\alpha _{3}\beta _{1}\beta _{2}\beta _{3}\beta _{4}-2\alpha _{1}\alpha
_{4}\beta _{1}\beta _{2}\beta _{3}\beta _{4}-4\alpha _{1}\alpha _{3}\beta
_{2}^{2}\beta _{3}\beta _{4}+4\alpha _{2}\alpha _{4}\beta _{2}^{2}\beta
_{3}\beta _{4}-\alpha _{2}\alpha _{3}\beta _{1}^{2}\beta _{4}^{2}+\alpha
_{1}\alpha _{4}\beta _{1}^{2}\beta _{4}^{2}+4\alpha _{1}\alpha _{3}\beta
_{1}\beta _{2}\beta _{4}^{2}-4\alpha _{2}\alpha _{4}\beta _{1}\beta
_{2}\beta _{4}^{2}-4\alpha _{2}\alpha _{3}\beta _{2}^{2}\beta
_{4}^{2}+4\alpha _{1}\alpha _{4}\beta _{2}^{2}\beta _{4}^{2})\det (\alpha
)\det^{2}(\gamma )\det^{2}(\delta )$.

The proofs of the properties for $F_{i}$ are similar to the ones of class $%
|\pi _{4}\rangle $. Next we prove if $F_{3}F_{4}=0$ then $F_{9}=F_{10}\neq 0$%
. Assume ${\small F_{3}=0}$. Then $\beta _{2}=0$. Then $F_{9}=$ $%
F_{10}=1/36(\beta _{1}^{2}\beta _{4}^{2}(\alpha _{1}\alpha _{4}-\alpha
_{2}\alpha _{3})\neq 0$ because $\alpha $ and $\beta $\ are invertible. As
well, we can show if ${\small F_{4}=0}$ then ${\small F_{9}=F_{10}\neq 0}$.

11. Class $|\sigma _{4}\rangle $

$\bigskip $(1). $IV=1/3\ast T$.

$D_{2}=-\frac{1}{36}(2\alpha _{1}\alpha _{3}\beta _{1}\beta _{3}+2\alpha
_{2}\alpha _{4}\beta _{1}\beta _{3}+\alpha _{2}\alpha _{3}\beta _{2}\beta
_{3}+\alpha _{1}\alpha _{4}\beta _{2}\beta _{3}+\alpha _{2}\alpha _{3}\beta
_{1}\beta _{4}+\alpha _{1}\alpha _{4}\beta _{1}\beta _{4}+2\alpha _{2}\alpha
_{4}\beta _{2}\beta _{4})\det (\alpha )\det (\beta )\det^{2}(\gamma
)\det^{2}(\delta \det (\alpha )\det (\alpha )).$

(2). $|F_{1}|+|F_{2}|\neq 0,|F_{3}|+|F_{4}|\neq 0$, if ${\small F_{1}F_{2}=0}
$ and ${\small F}_{3}{\small F}_{4}={\small 0}$ then ${\small F}_{9}{\small %
=0\wedge F}_{10}{\small =0}$.

Proof.

$F_{1}=(1/9)\alpha _{2}^{4}P$, $F_{2}=(1/9)\alpha _{4}^{4}P$, $%
F_{3}=(1/9)\beta _{1}^{4}Q$, $F_{4}=(1/9)\beta _{3}^{4}Q$,

$F_{9}=(1/9)(\alpha _{2}\alpha _{3}\beta _{1}\beta _{3}-\alpha _{1}\alpha
_{4}\beta _{1}\beta _{3}-\alpha _{2}\alpha _{4}\beta _{2}\beta _{3}+\alpha
_{2}\alpha _{4}\beta _{1}\beta _{4})^{2}\det^{2}(\delta )\det^{2}(\gamma )$,

$F_{10}=(1/9)(\alpha _{2}\alpha _{3}\beta _{1}\beta _{3}-\alpha _{1}\alpha
_{4}\beta _{1}\beta _{3}+\alpha _{2}\alpha _{4}\beta _{2}\beta _{3}-\alpha
_{2}\alpha _{4}\beta _{1}\beta _{4})^{2}\det^{2}(\delta )\det^{2}(\gamma ).$

12. Class $|\rho _{4}\rangle $

(1). $IV=1/3\ast T$. $D_{2}=-\frac{1}{36}(2\alpha _{1}\alpha _{3}\beta
_{1}\beta _{3}+\alpha _{2}\alpha _{3}\beta _{2}\beta _{3}+\alpha _{1}\alpha
_{4}\beta _{2}\beta _{3}+\alpha _{2}\alpha _{3}\beta _{1}\beta _{4}+\alpha
_{1}\alpha _{4}\beta _{1}\beta _{4}+2\alpha _{2}\alpha _{4}\beta _{2}\beta
_{4})\det (\alpha )\det (\beta )\det^{2}(\gamma )\det^{2}(\delta )$.

(2). $F_{5}=\frac{1}{9}\gamma _{2}^{4}R$, $F_{6}=\frac{1}{9}\gamma _{4}^{4}R$%
, $F_{7}=\frac{1}{9}\delta _{1}^{4}S$, $F_{8}=\frac{1}{9}\delta _{3}^{4}S$.

13. Class $|\xi _{4}\rangle $

(1). $IV=-1/3\ast T$.

$D_{3}=-\frac{1}{36}(2\alpha _{1}\alpha _{3}\delta _{1}\delta _{3}+2\alpha
_{2}\alpha _{4}\delta _{1}\delta _{3}+\alpha _{2}\alpha _{3}\delta
_{2}\delta _{3}+\alpha _{1}\alpha _{4}\delta _{2}\delta _{3}+\alpha
_{2}\alpha _{3}\delta _{1}\delta _{4}+\alpha _{1}\alpha _{4}\delta
_{1}\delta _{4}+2\alpha _{2}\alpha _{4}\delta _{2}\delta _{4})\det (\alpha
)\det^{2}(\beta )\det^{2}(\gamma )\det (\delta )$,

(2). $|F_{1}|+|F_{2}|\neq 0,|F_{7}|+|F_{8}|\neq 0,$ $\#2$.

Proof of (2):

$F_{1}=\frac{1}{9}\alpha _{2}^{4}P$, $F_{2}=\frac{1}{9}\alpha _{4}^{4}P$, $%
F_{7}=\frac{1}{9}\delta _{1}^{4}S$, $F_{8}=\frac{1}{9}\delta _{3}^{4}S$,

$F_{9}$ $=\frac{1}{36}(4\alpha _{2}^{2}\alpha _{3}\alpha _{4}\beta _{1}\beta
_{3}-4\alpha _{1}\alpha _{2}\alpha _{4}^{2}\beta _{1}\beta _{3}-\alpha
_{2}^{2}\alpha _{3}^{2}\beta _{2}\beta _{3}+2\alpha _{1}\alpha _{2}\alpha
_{3}\alpha _{4}\beta _{2}\beta _{3}-\alpha _{1}^{2}\alpha _{4}^{2}\beta
_{2}\beta _{3}-4\alpha _{2}^{2}\alpha _{4}^{2}\beta _{2}\beta _{3}+\alpha
_{2}^{2}\alpha _{3}^{2}\beta _{1}\beta _{4}-2\alpha _{1}\alpha _{2}\alpha
_{3}\alpha _{4}\beta _{1}\beta _{4}+\alpha _{1}^{2}\alpha _{4}^{2}\beta
_{1}\beta _{4}+4\alpha _{2}^{2}\alpha _{4}^{2}\beta _{1}\beta _{4}-4\alpha
_{2}^{2}\alpha _{3}\alpha _{4}\beta _{2}\beta _{4}+4\alpha _{1}\alpha
_{2}\alpha _{4}^{2}\beta _{2}\beta _{4})\det (\beta )\det^{2}(\gamma
)\det^{2}(\delta )$,

$F_{10}=\frac{1}{36}(-4\alpha _{2}^{2}\alpha _{3}\alpha _{4}\beta _{1}\beta
_{3}+4\alpha _{1}\alpha _{2}\alpha _{4}^{2}\beta _{1}\beta _{3}-\alpha
_{2}^{2}\alpha _{3}^{2}\beta _{2}\beta _{3}+2\alpha _{1}\alpha _{2}\alpha
_{3}\alpha _{4}\beta _{2}\beta _{3}-\alpha _{1}^{2}\alpha _{4}^{2}\beta
_{2}\beta _{3}-4\alpha _{2}^{2}\alpha _{4}^{2}\beta _{2}\beta _{3}+\alpha
_{2}^{2}\alpha _{3}^{2}\beta _{1}\beta _{4}-2\alpha _{1}\alpha _{2}\alpha
_{3}\alpha _{4}\beta _{1}\beta _{4}+\alpha _{1}^{2}\alpha _{4}^{2}\beta
_{1}\beta _{4}+4\alpha _{2}^{2}\alpha _{4}^{2}\beta _{1}\beta _{4}+4\alpha
_{2}^{2}\alpha _{3}\alpha _{4}\beta _{2}\beta _{4}-4\alpha _{1}\alpha
_{2}\alpha _{4}^{2}\beta _{2}\beta _{4})\det (\beta )\det^{2}(\gamma
)\det^{2}(\delta )$.

The rest proofs are similar to the ones of $|\pi _{4}\rangle $.

14. Class $|\epsilon _{4}\rangle $

(1). $IV=-1/3\ast T$.

$D_{3}=-\frac{1}{36}(2\alpha _{1}\alpha _{3}\delta _{1}\delta _{3}+\alpha
_{2}\alpha _{3}\delta _{2}\delta _{3}+\alpha _{1}\alpha _{4}\delta
_{2}\delta _{3}+\alpha _{2}\alpha _{3}\delta _{1}\delta _{4}+\alpha
_{1}\alpha _{4}\delta _{1}\delta _{4}+2\alpha _{2}\alpha _{4}\delta
_{2}\delta _{4})\det (\alpha )\det^{2}(\beta )\det^{2}(\gamma )\det (\delta
) $.

(2). $|F_{3}|+|F_{4}|\neq 0,|F_{5}|+|F_{6}|\neq 0,$ $\#3$.

Proof.

$F_{3}=\frac{1}{9}\beta _{1}^{4}Q$, $F_{4}=$ $\frac{1}{9}\beta _{3}^{4}Q$, $%
F_{5}=$ $\frac{1}{9}\gamma _{2}^{4}R$, $F_{6}=$ $\frac{1}{9}\gamma _{4}^{4}R$%
,

$F_{9}=\frac{1}{36}(-4\alpha _{2}\alpha _{3}\beta _{1}^{2}\beta
_{3}^{2}+4\alpha _{1}\alpha _{4}\beta _{1}^{2}\beta _{3}^{2}-4\alpha
_{1}\alpha _{3}\beta _{1}\beta _{2}\beta _{3}^{2}+4\alpha _{2}\alpha
_{4}\beta _{1}\beta _{2}\beta _{3}^{2}-\alpha _{2}\alpha _{3}\beta
_{2}^{2}\beta _{3}^{2}+\alpha _{1}\alpha _{4}\beta _{2}^{2}\beta
_{3}^{2}+4\alpha _{1}\alpha _{3}\beta _{1}^{2}\beta _{3}\beta _{4}-4\alpha
_{2}\alpha _{4}\beta _{1}^{2}\beta _{3}\beta _{4}+2\alpha _{2}\alpha
_{3}\beta _{1}\beta _{2}\beta _{3}\beta _{4}-2\alpha _{1}\alpha _{4}\beta
_{1}\beta _{2}\beta _{3}\beta _{4}-\alpha _{2}\alpha _{3}\beta _{1}^{2}\beta
_{4}^{2}+\alpha _{1}\alpha _{4}\beta _{1}^{2}\beta _{4}^{2})\det (\alpha
)\det^{2}(\gamma )\det^{2}(\delta )$,

$F_{10}=-\frac{1}{36}(4\alpha _{2}\alpha _{3}\beta _{1}^{2}\beta
_{3}^{2}-4\alpha _{1}\alpha _{4}\beta _{1}^{2}\beta _{3}^{2}-4\alpha
_{1}\alpha _{3}\beta _{1}\beta _{2}\beta _{3}^{2}+4\alpha _{2}\alpha
_{4}\beta _{1}\beta _{2}\beta _{3}^{2}+\alpha _{2}\alpha _{3}\beta
_{2}^{2}\beta _{3}^{2}-\alpha _{1}\alpha _{4}\beta _{2}^{2}\beta
_{3}^{2}+4\alpha _{1}\alpha _{3}\beta _{1}^{2}\beta _{3}\beta _{4}-4\alpha
_{2}\alpha _{4}\beta _{1}^{2}\beta _{3}\beta _{4}-2\alpha _{2}\alpha
_{3}\beta _{1}\beta _{2}\beta _{3}\beta _{4}+2\alpha _{1}\alpha _{4}\beta
_{1}\beta _{2}\beta _{3}\beta _{4}+\alpha _{2}\alpha _{3}\beta _{1}^{2}\beta
_{4}^{2}-\alpha _{1}\alpha _{4}\beta _{1}^{2}\beta _{4}^{2})\det (\alpha
)\det^{2}(\gamma )\det^{2}(\delta )$.

The rest proofs of the properties for $F_{i}$ are similar to the ones of $%
|\theta _{4}\rangle $.

15. Class $|\chi _{4}\rangle $

(1). $D_{1}=-(1/18)(\alpha _{1}\alpha _{3}-\alpha _{2}\alpha _{4})\gamma
_{2}\gamma _{4}\det (\alpha )\det^{2}(\beta )\det (\gamma )\det^{2}(\delta )$%
,

\ $D_{2}=(1/36)(\alpha _{2}\alpha _{3}+\alpha _{1}\alpha _{4})(\beta
_{2}\beta _{3}+\beta _{1}\beta _{4})\det (\alpha )\det (\beta
)\det^{2}(\gamma )\det^{2}(\delta )$,

$\ \ D_{3}=(1/18)(\alpha _{1}\alpha _{3}+\alpha _{2}\alpha _{4})\delta
_{1}\delta _{3}\det (\alpha )\det^{2}(\beta )\det^{2}(\gamma )\det (\delta )$%
.

(2). ${\small |F_{5}|+|F_{6}|\neq 0}$, ${\small |F_{7}|+|F_{8}|\neq 0}$, $%
\#0 $.

Proof of (2):

$F_{1}=\allowbreak \frac{1}{9}\alpha _{1}^{2}\alpha _{2}^{2}P$, $F_{2}=\frac{%
1}{9}\alpha _{3}^{2}\alpha _{4}^{2}P$, $F_{3}=\frac{1}{9}\beta _{1}^{2}\beta
_{2}^{2}Q$, $F_{4}=\allowbreak \frac{1}{9}\beta _{3}^{2}\beta _{4}^{2}Q$, $%
F_{5}=\frac{1}{9}(\gamma _{1}-\gamma _{2})\gamma _{2}^{2}(\gamma _{1}+\gamma
_{2})R$, $F_{6}=\allowbreak \frac{1}{9}(\gamma _{3}-\gamma _{4})\gamma
_{4}^{2}(\gamma _{3}+\gamma _{4})R$, $F_{7}=\allowbreak \frac{1}{9}\delta
_{1}^{2}(\delta _{1}^{2}+\delta _{2}^{2})S$, $F_{8}=\frac{1}{9}\delta
_{3}^{2}(\delta _{3}^{2}+\delta _{4}^{2})S$,

$F_{9}=\allowbreak \frac{1}{9}\left( \alpha _{1}\beta _{1}\alpha _{4}\beta
_{4}-\beta _{3}\alpha _{3}\beta _{2}\alpha _{2}\right) ^{2}\det^{2}(\delta
)\det^{2}(\gamma )$,

$F_{10}=\frac{1}{9}\left( -\alpha _{1}\beta _{3}\alpha _{4}\beta _{2}+\beta
_{1}\alpha _{3}\beta _{4}\alpha _{2}\right) ^{2}\allowbreak \det^{2}(\delta
)\det^{2}(\gamma )$.

Let us show ${\small |F_{5}|+|F_{6}|\neq 0}$. Case 1, $\gamma _{2}=0$. Then $%
\gamma _{4}\neq 0$ because $\gamma $ is invertible. Next we show $\gamma
_{3}^{2}\neq \gamma _{4}^{2}$. Otherwise,\ $\gamma _{4}=0$ when $\gamma
_{3}=0$. This contradicts that $\gamma $ is invertible. Case 2, $\gamma
_{4}=0$. Similarly, we can show $\gamma _{2}\neq 0$ and $\gamma _{1}^{2}\neq
\gamma _{2}^{2}$. Case 3, $\gamma _{1}^{2}=\gamma _{2}^{2}$. Then $\gamma
_{3}^{2}\neq \gamma _{4}^{2}$. Otherwise, $\gamma _{1}^{2}\gamma
_{4}^{2}=\gamma _{2}^{2}\gamma _{3}^{2}$. Since $\gamma _{1}\gamma
_{4}-\gamma _{2}\gamma _{3}\neq 0$, $\gamma _{1}\gamma _{4}+\gamma
_{2}\gamma _{3}=0$. Thus, $\det (\gamma )=-2\gamma _{2}\gamma _{3}$.
Clearly, for an invertible $\gamma $ in which $\gamma _{2}\gamma _{3}=0$, $%
\det (\gamma )=0$. This a paradox. Case 4, $\gamma _{3}^{2}=\gamma _{4}^{2}$%
. Similarly, $\gamma _{1}^{2}\neq \gamma _{2}^{2}$.

The rest proof is similar to the one of $|GHZ\rangle $.

16. Class $|\upsilon _{4}\rangle $

(1). $D_{2}=\frac{1}{16}\alpha _{1}\alpha _{3}\beta _{1}\beta _{3}\det
(\alpha )\det (\beta )\det^{2}(\gamma )\det^{2}(\delta )$,

$\ \ \ D_{3}=-\frac{1}{16}\alpha _{1}\alpha _{3}\delta _{2}\delta _{4}\det
(\alpha )\det^{2}(\beta )\det^{2}(\gamma )\det (\delta )$.

(2). $\#1$.

$F_{1}=\frac{1}{16}\alpha _{1}^{4}P$, $F_{2}=\frac{1}{16}\alpha _{3}^{4}P$, $%
F_{3}=\frac{1}{16}\beta _{1}^{4}Q$, $F_{4}=\frac{1}{16}\beta _{3}^{4}Q$, $%
F_{7}=\frac{1}{16}\delta _{2}^{4}S$, $F_{8}=\frac{1}{16}\delta _{4}^{4}S$,

$F_{9}=\frac{1}{16}(\alpha _{2}\alpha _{3}\beta _{1}\beta _{3}-\alpha
_{1}\alpha _{4}\beta _{1}\beta _{3}+\alpha _{1}\alpha _{3}\beta _{2}\beta
_{3}-\alpha _{1}\alpha _{3}\beta _{1}\beta _{4})^{2}\det^{2}(\gamma
)\det^{2}(\delta )$,

$F_{10}=\frac{1}{16}(-\alpha _{2}\alpha _{3}\beta _{1}\beta _{3}+\alpha
_{1}\alpha _{4}\beta _{1}\beta _{3}+\alpha _{1}\alpha _{3}\beta _{2}\beta
_{3}-\alpha _{1}\alpha _{3}\beta _{1}\beta _{4})^{2}\det^{2}(\gamma
)\det^{2}(\delta )$.

\bigskip 17. Class $|\varpi _{4}\rangle $

(1). $D_{1}=-\frac{1}{16}\alpha _{1}\alpha _{3}\gamma _{1}\gamma _{3}\det
(\alpha )\det^{2}(\beta )\det (\gamma )\det^{2}(\delta )$,

$\ \ \ D_{3}=\frac{1}{16}\alpha _{1}\alpha _{3}\delta _{1}\delta _{3}\det
(\alpha )\det^{2}(\beta )\det^{2}(\gamma )\det (\delta ).$

(2). $F_{1}=\frac{1}{16}\alpha _{1}^{4}P$, $F_{2}=\frac{1}{16}\alpha
_{3}^{4}P$, $F_{5}=\frac{1}{16}\gamma _{1}^{4}R$, $F_{6}=\frac{1}{16}\gamma
_{3}^{4}R$, $F_{7}=\frac{1}{16}\delta _{1}^{4}S$, $F_{8}=\frac{1}{16}\delta
_{3}^{4}S$, $F_{9}=\frac{1}{16}\alpha _{1}^{2}\alpha _{3}^{2}P$, $%
F_{10}=F_{9}$.

18. Class $|\psi _{4}\rangle $

(1). $F>0$, $F_{9}=F_{10}$, if $F_{i}=F_{j}=F_{k}=0$, where $1\leq i<j<k\leq
4,$ then $F_{9}\neq 0$.

(2). $D_{1}=(-1/16)(\alpha _{2}\alpha _{3}+\alpha _{1}\alpha _{4})(\gamma
_{2}\gamma _{3}+\gamma _{1}\gamma _{4})\det (\alpha )\det^{2}(\beta )\det
(\gamma )\det^{2}(\delta )$.

\noindent Proof of (1):

The values of $F_{1}$ to $F_{8}$ are the same as the ones of $F_{1}$ to $%
F_{8}$ of class $|GHZ\rangle $, respectively.

$F_{9}=(1/16)(\alpha _{2}^{2}\alpha _{3}^{2}\beta _{2}^{2}\beta
_{3}^{2}+2\alpha _{1}\alpha _{2}\alpha _{3}\alpha _{4}\beta _{2}^{2}\beta
_{3}^{2}+\alpha _{1}^{2}\alpha _{4}^{2}\beta _{2}^{2}\beta _{3}^{2}+2\alpha
_{2}^{2}\alpha _{3}^{2}\beta _{1}\beta _{2}\beta _{3}\beta _{4}-12\alpha
_{1}\alpha _{2}\alpha _{3}\alpha _{4}\beta _{1}\beta _{2}\beta _{3}\beta
_{4}+2\alpha _{1}^{2}\alpha _{4}^{2}\beta _{1}\beta _{2}\beta _{3}\beta
_{4}+\alpha _{2}^{2}\alpha _{3}^{2}\beta _{1}^{2}\beta _{4}^{2}+2\alpha
_{1}\alpha _{2}\alpha _{3}\alpha _{4}\beta _{1}^{2}\beta _{4}^{2}+\alpha
_{1}^{2}\alpha _{4}^{2}\beta _{1}^{2}\beta _{4}^{2})\det^{2}(\gamma
)\det^{2}(\beta )$,

$F_{10}=F_{9}.$

Let us prove that if $F_{1}=F_{3}=F_{4}=0$ then $F_{9}\neq 0$. The proofs
for other cases are similar.

Assume that $F_{1}=0$. Then there are two cases: case 1, $\alpha _{1}=0$ and
case 2, $\alpha _{2}=0$.

Case 1. $\alpha _{1}=0$. In this case, $\alpha _{2}\alpha _{3}\neq 0$. Thus,

$F_{9}=(1/16)\allowbreak \alpha _{2}^{2}\alpha _{3}^{2}\left( \beta
_{1}\beta _{4}+\beta _{2}\beta _{3}\right) ^{2}\det^{2}(\gamma
)\det^{2}(\beta )$.

Since $F_{3}=F_{4}=0$, there are two cases.

Case 1.1 $\beta _{1}=\beta _{4}=0$. Then $\beta _{2}\beta _{3}\neq 0$.

Case 1.2. $\beta _{2}=\beta _{3}=0$. Then $\beta _{1}\beta _{4}\neq 0$.

In either case, it is straightforward that $F_{9}\neq 0$.

Case 2. $\alpha _{2}=0$. In this case, $\alpha _{1}\alpha _{4}\neq 0$. Thus,

$F_{9}=(1/16)\allowbreak \alpha _{1}^{2}\alpha _{4}^{2}\left( \beta
_{1}\beta _{4}+\beta _{2}\beta _{3}\right) ^{2}\det^{2}(\gamma
)\det^{2}(\beta )$.

As discussed above, when $F_{3}=F_{4}=0$, $F_{9}\neq 0$.

19. Class $|\phi _{4}\rangle $

(1). $F>0.$

\noindent\ \ \ (2). $D_{2}=(1/16)(\alpha _{2}\alpha _{3}+\alpha _{1}\alpha
_{4})(\beta _{2}\beta _{3}+\beta _{1}\beta _{4})\det (\alpha )\det (\beta
)\det^{2}(\gamma )\det^{2}(\delta )$.

(3). $\ \#0$.

Proof.

$F_{1}$ to $F_{10}$ are the same as $F_{1}$ to $F_{10}$ of class $%
|GHZ\rangle $. The next argument is the same as the one of class $%
|GHZ\rangle $.

20. Class $|\mu _{4}\rangle $

(1). $F_{9}=F_{10}$, if $F_{i}=F_{j}=F_{k}=0$, where $1\leq i<j<k\leq 4,$
then $F_{9}\neq 0$.

(2). $D_{3}=(1/16)(\alpha _{2}\alpha _{3}+\alpha _{1}\alpha _{4})(\delta
_{2}\delta _{3}+\delta _{1}\delta _{4})\det (\alpha )\det^{2}(\beta
)\det^{2}(\gamma )\det (\delta )$.

\noindent $F_{1}$ to $F_{10}$ are the same as the ones of class $|\psi
_{4}\rangle .$ The argument is the same as the one for class $|\psi
_{4}\rangle $.

21. Class $|\varphi _{4}\rangle $

(1) $D_{1}=\frac{1}{9}\alpha _{1}\alpha _{3}\gamma _{2}\gamma _{4}\det
(\alpha )\det^{2}(\beta )\det (\gamma )\det^{2}(\delta )$.

(2). $F_{1}=\frac{1}{9}\alpha _{1}^{4}P$, $F_{2}=\frac{1}{9}\alpha _{3}^{4}P 
$, $F_{5}=\frac{1}{9}\gamma _{2}^{4}R$, $F_{6}=\frac{1}{9}\gamma _{4}^{4}R$, 
$F_{9}=\frac{1}{9}\alpha _{1}^{2}\alpha _{3}^{2}P$, $F_{10}=F_{9}$.

\bigskip 22. Class $|\zeta _{4}\rangle $

(1) $D_{2}=-\frac{1}{9}\alpha _{2}\alpha _{4}\beta _{1}\beta _{3}\det
(\alpha )\det (\beta )\det^{2}(\gamma )\det^{2}(\delta )$.

(2). $\#1$.

The values of $F_{i}$ are the same as the ones of $F_{i}$ for class $|\sigma
_{4}\rangle $.

23. Class $|\vartheta _{4}\rangle $

(1) $D_{3}=-\frac{1}{9}\alpha _{1}\alpha _{3}\delta _{2}\delta _{4}\det
(\alpha )\det^{2}(\beta )\det^{2}(\gamma )\det (\delta )$.

(2). $F_{1}=\frac{1}{9}\alpha _{1}^{4}P$, $F_{2}=\frac{1}{9}\alpha _{3}^{4}P 
$, $F_{7}=\frac{1}{9}\delta _{2}^{4}S$, $F_{8}=\frac{1}{9}\delta _{4}^{4}S$, 
$F_{9}=\frac{1}{9}\alpha _{1}^{2}\alpha _{3}^{2}P$, $F_{10}=F_{9}$.

24. Class $|\tau _{4}\rangle $

$F_{3}=\frac{1}{9}\beta _{1}^{4}Q$, $F_{4}=\frac{1}{9}\beta _{3}^{4}Q$, $%
F_{5}=\frac{1}{9}\gamma _{2}^{4}R$, $F_{6}=\frac{1}{9}\gamma _{4}^{4}R$, $%
F_{9}=\frac{1}{9}\beta _{1}^{2}\beta _{3}^{2}Q$, $F_{10}=F_{9}$.

25. Class $|\varrho _{4}\rangle $

$F_{3}=\frac{1}{9}\beta _{1}^{4}Q$, $F_{4}=\frac{1}{9}\beta _{3}^{4}Q$, $%
F_{7}=\frac{1}{9}\delta _{2}^{4}S$, $F_{8}=\frac{1}{9}\delta _{4}^{4}S$, $%
F_{9}=\frac{1}{9}\beta _{1}^{2}\beta _{3}^{2}Q$, $F_{10}=F_{9}$.

26. Class $|\iota _{4}\rangle $

$F_{5}=\gamma _{1}^{4}R/9$, $F_{6}=\gamma _{3}^{4}R/9$, $F_{7}=\delta
_{1}^{4}S/9$, $F_{8}=\delta _{3}^{4}S/9$.

27. Class $|\omega _{4}\rangle $

\noindent $F_{3}=\beta _{2}^{4}Q/16$, $F_{4}=\beta _{4}^{4}Q/16$, $%
F_{5}=\gamma _{1}^{4}R/16$, $F_{6}=\gamma _{3}^{4}R/16$, $F_{7}=\delta
_{1}^{4}S/16$, $F_{8}=\delta _{3}^{4}S/16$, $F_{9}=F_{10}=\beta
_{2}^{2}\beta _{4}^{2}Q/16$.

\textbf{Appendix B: The number of the degenerated SLOCC classes for }$n$%
\textbf{-qubits}

Let $d(n)$ be the number of the degenerated SLOCC classes for $n$-qubits and 
$t(n)$ the number of the true entanglement SLOCC classes for $n$-qubits and $%
t(1)=1$.

1. \textbf{By computing }$d(5)$\textbf{\ demonstrate how to derive }$d(n)$

Case 1. Only 4-qubit true entanglement accompanied with a separable qubit

For example, $ABCD-E$, where $ABCD$ is truly entangled. Note that
four-qubits can truly be entangled in $t(4)$\ distinct ways.\ Clearly, there
are $\left( _{4}^{5}\right) t(4)$ distinct degenerated SLOCC classes. This
situation can be considered as that five balls are divided into two groups.
The first group contains exactly one ball and the second group contains
exactly 4 balls. Thus, there are $\frac{5!}{1!4!}$ different ways\cite%
{DeGroot}.\ Note that four balls correspond to four qubits. Therefore, for
the case, the number of the degenerated SLOCC classes can be rewritten as $%
\frac{5!}{1!4!}t(1)t(4)$. We can consider this case as a partition of 5: $%
1+4=5$.

Case 2. Only 3-qubit true entanglement accompanied with two separable qubits

For example, $ABC-D-E$, where $ABC$ is truly entangled. As indicated in \cite%
{Dur}, three qubits can truly be entangled in two inequivalent ways. It is
easy to see that there are\ $2\left( _{3}^{5}\right) $ distinct degenerated
SLOCC\ classes. Let us consider that five balls are divided into three
groups. The first two groups contain exactly one ball respectively and the
third group contains exactly three balls. Thus, there are $\frac{5!}{1!1!3!}$
different ways\cite{DeGroot}. Note that $ABC-D-E$ and $ABC-E-D$\ represent
the same class. Hence, for the case, the number of the degenerated SLOCC
classes can be rewritten as $\frac{5!}{1!1!3!}\ast t(1)t(1)t(3)\ast \frac{1}{%
2!}$. We can remember this case as a partition of 5: $1+1+3=5$.

Case 3. Only 2-qubit true entanglement accompanied with three separable
qubits

For example, $AB-C-D-E$, where $AB$ is a two-qubit $GHZ$ state. It is clear
that there are\ $\left( _{2}^{5}\right) $ distinct classes. We consider that
five balls are divided into four groups. The first three groups contain
exactly one ball respectively and the fourth group contains exactly two
balls. Thus, there are $\frac{5!}{1!1!1!2!}$ different ways\cite{DeGroot}.
Note that $AB-X-Y-Z$, where $X$, $Y$ and $Z$ is any one of $3!$ permutations
of $C$, $D$ and $E$, represent the same class. Then, for the case, the
number of the degenerated SLOCC classes can be rewritten as $\frac{5!}{%
1!1!1!2!}\ast t(1)t(1)t(1)t(2)\ast \frac{1}{3!}$. Let us consider this case
as a partition of 5: $1+1+1+2=5$.

Case 4. Two $GHZ$ pairs accompanied with a separable qubit

For example, $AB-CD-E$, where $AB$ and $CD$ are two-qubit $GHZ$ states. For
this case, there are $15$ degenerated SLOCC classes. Note that $AB-CD-E$\
and $CD-AB-E$\ represent the same class. Similarly, for the case, the number
of the degenerated SLOCC classes can be rewritten as $\frac{5!}{1!2!2!}\ast
t(1)t(2)t(2)\ast \frac{1}{2!}$. We can consider this case as a partition of
5: $1+2+2=5$.

Case 5. two-qubit $GHZ\otimes $ three-qubit $GHZ$ and two-qubit $GHZ\otimes $
three-qubit $W$

For example, $AB-CDE$.\ We can list the entanglement ways as follows.

$AB-CDE$; $AC-BDE$; $AD-BCE$; $AE-BCD$; $BC-ADE$; $BD-ACE$; $BE-ACD$; $%
CD-ABE $; $CE-ABD$; $DE-ABC$.

Hence, there are $2\left( _{2}^{5}\right) $ degenerated SLOCC classes.
Similarly, for the case, the number of the degenerated SLOCC classes can be
rewritten as $\frac{5!}{2!3!}\ast t(2)t(3)$. We can remember this case as a
partition of 5: $2+3=5$.

Case 6. A product state: $A-B-C-D-E$

It is trivial that $\frac{5!}{1!1!1!1!1!}\ast t(1)t(1)t(1)t(1)t(1)\ast \frac{%
1}{5!}=1$. Let us consider this case as a partition of 5: $1+1+1+1+1=5$.

In total, there are $5\ast t(4)+66$ degenerated SLOCC classes for
five-qubits.\ 

2. \textbf{The exact recursive formula of }$d(n)$

For $n-$qubits, we consider the degenerated entanglement way $r_{1}\otimes
r_{2}\otimes ...\otimes r_{k}$, where $k\geq 2$ and the $r_{i}$ qubits are
truly entangled, $i=1,2,...,k$. As indicated in \cite{DeGroot}, there are $%
\frac{n!}{r_{1}!r_{2}!..r_{k}!}$ different ways to divide $n$ balls into $k $
groups such that the $j$th group contains exactly $r_{j}$ balls, where $%
r_{1}+r_{2}+...+r_{k}=n$. Note that $r_{i}$ qubits can truly be entangled in 
$t(r_{i})$ distinct ways. Let $s_{j}$ be the number of the concurrences of $%
r_{i_{j}}$ in $r_{1},r_{2},...,r_{k}$. Thus, for the situation, there are $%
\frac{n!}{r_{1}!r_{2}!..r_{k}!}t(r_{1})t(r_{2})...t(r_{k})\frac{1}{%
s_{1}!s_{2}!...s_{l}!}$ degenerated SLOCC classes. In total, $d(n)=$ $\sum 
\frac{n!}{r_{1}!r_{2}!..r_{k}!}t(r_{1})t(r_{2})...t(r_{k})\frac{1}{%
s_{1}!s_{2}!...s_{l}!}$, where the summation is extended over all the
following Euler's partitions of $n$: $r_{1}+r_{2}+...+r_{k}=n$ in which $%
k\geq 2$ and $1\leq r_{1}\leq r_{2}\leq ...\leq r_{k}<n$.

3. \textbf{Compute }$d(4)$\textbf{\ using the recursive formula}

For four-qubits, the following are the partitions of 4.

$1+3$; $1+1+2$; $2+2$; $1+1+1+1.$

Case 1. For the partition $1+3$, there are $\frac{4!}{1!3!}t(1)t(3)=8$
degenerated SLOCC classes. They are $A-BCD$, $B-ACD$, $C-ABD$, $D-ABC$. Note
that three qubits can truly be entangled in two inequivalent ways.\ \ 

Case 2. For the partition $1+1+2$, there are $\frac{4!}{1!1!2!}t(1)t(1)t(2)%
\frac{1}{2!}=6$ degenerated SLOCC classes. They are $A-B-CD$, $A-C-BD$, $%
A-D-BC$, $B-C-AD$, $B-D-AC$, $C-D-AB$.

Case 3. For the partition $2+2$, there are $\frac{4!}{2!2!}t(2)t(2)\frac{1}{%
2!}=3$ degenerated SLOCC classes. They are $AB-CD$, $AC-BD$, $AD-BC$.

Case 4. For the partition $1+1+1+1$, \ this case is a product state. It is
trivial that $\frac{4!}{1!1!1!1!}\ast t(1)t(1)t(1)t(1)\ast \frac{1}{4!}=1$.

Totally, there are 18 degenerated SLOCC classes. See \cite{Moor2}.

\ \ \ \ \

\end{document}